\documentclass[nofootinbib,twocolumn,prd,preprintnumbers,superscriptaddress]{revtex4-1}

\usepackage{amsfonts}
\usepackage{graphicx}
\usepackage[colorlinks=true,linkcolor=blue,citecolor=red,urlcolor=blue]{hyperref}
\usepackage{color}
\usepackage{amsmath}
\usepackage{cases}
\usepackage{braket}
\usepackage[T1]{fontenc}
\usepackage{mathrsfs}
\usepackage{bm}

\begin{document}
\title{Role of spatial curvature in the primordial gravitational wave power spectrum}

\author{Rocco D'Agostino}
\email{rocco.dagostino@unina.it}
\affiliation{Scuola Superiore Meridionale, Largo San Marcellino 10, 80138 Napoli, Italy}
\affiliation{Istituto Nazionale di Fisica Nucleare (INFN), Sezione di Napoli, Via Cinthia 21,  80126 Napoli, Italy}

\author{Matteo Califano}
\email{matteo.califano@unina.it}
\affiliation{Scuola Superiore Meridionale, Largo San Marcellino 10, 80138 Napoli, Italy}
\affiliation{Istituto Nazionale di Fisica Nucleare (INFN), Sezione di Napoli, Via Cinthia 21, 80126 Napoli, Italy}

\author{Nicola Menadeo}
\email{nicola.menadeo-ssm@unina.it}
\affiliation{Scuola Superiore Meridionale, Largo San Marcellino 10, 80138 Napoli, Italy}
\affiliation{Istituto Nazionale di Fisica Nucleare (INFN), Sezione di Napoli, Via Cinthia 21, 80126 Napoli, Italy}

\author{Daniele Vernieri}
\email{daniele.vernieri@unina.it}
\affiliation{Dipartimento di Fisica ``E. Pancini'', Universit\`a di Napoli ``Federico II'', Via Cinthia 21, 80126 Napoli, Italy}
\affiliation{Scuola Superiore Meridionale, Largo San Marcellino 10, 80138 Napoli, Italy}
\affiliation{Istituto Nazionale di Fisica Nucleare (INFN), Sezione di Napoli, Via Cinthia 21, 80126 Napoli, Italy}

\begin{abstract}
This paper investigates the effects of nonvanishing spatial curvature on the propagation of primordial gravitational waves produced during inflation.
In particular, we consider tensor perturbations over a homogeneous and isotropic background, and describe the propagation of gravitational waves in the de Sitter phase with spatially curved geometries.
We thus derive the expression of the primordial power spectrum at the horizon crossing, in the case of open and closed universes.
Then, we analyze how tensor modes propagate in the post-inflationary era, showing the evolution of transfer functions in the radiation and matter epochs, as well as the matching conditions in the intermediate regime. To account for the intrinsic nature of different relativistic species, we also explore the corrections to the standard behavior of the radiation energy density. For this purpose, we introduce the effective number of degrees of freedom of relativistic particles contributing to the primordial energy and entropy densities. Under the subhorizon approximation, we obtain the spectral energy density of relic gravitational waves in terms of the curvature density parameter. Finally, we discuss the capability of present and future experiments to detect the primordial gravitational wave signal at different frequency regimes.
\end{abstract}

\maketitle

\preprint{ET-0162A-23}

\section{Introduction}
 The success of the standard cosmological model has been pushed to unprecedented precision by the most recent measurements of the cosmic microwave background (CMB) anisotropies obtained by the Planck collaboration \cite{Planck:2018vyg}. The emerging picture based on the theory of general relativity indicates that the cosmic fluid is mainly composed of cold dark matter (CDM) and dark energy, under the simplest form of the cosmological constant $(\Lambda)$, which accounts for the accelerated expansion of the Universe observed at late times \cite{SupernovaSearchTeam:1998fmf,SupernovaCosmologyProject:1998vns,Carroll:2000fy,Peebles:2002gy,DAgostino:2022fcx}\footnote{Many alternative models have been proposed over the years to explain the dark energy effects, including modifications of the gravity sector \cite{Sotiriou:2008rp,Linder:2010py,DAgostino:2018ngy,BeltranJimenez:2019tme,Capozziello:2022rac}, unified cosmic fluids \cite{Kamenshchik:2001cp,Scherrer:2004au,Capozziello:2017buj} or scenarios based on the holographic principle \cite{Li:2004rb,DAgostino:2019wko,Saridakis:2020zol}. For comprehensive reviews, see e.g. Refs.~\cite{Clifton:2011jh,Capozziello:2019cav} and references therein.}. Within such a framework, spacetime is well approximated by the Friedmann-Lema\^itre-Robertson-Walker (FLRW) metric with vanishing spatial curvature.  
The nearly homogeneous and flat universe observed today can be explained by invoking the theory of cosmic inflation \cite{Starobinsky:1980te,Guth:1980zm,Linde:1981mu,DAgostino:2021vvv}, which provides the generation mechanism for density perturbations in the primordial Universe. According to this scenario, the tiny fluctuations in the CMB temperature represent the relic of quantum fluctuations in the earliest stages of the Universe's evolution \cite{Mukhanov:1990me}. The exponential expansion of spacetime characterizing the inflationary epoch leads to a spatially (almost) flat FLRW geometry, regardless of the initial conditions. However, albeit negligibly small at the present time, the spatial curvature effects may have played an important role before the onset of inflation.

Despite the predictive power of the flat $\Lambda$CDM model, recent findings do not exclude the possibility to deviate from zero spatial curvature. Indeed, the latest CMB Planck results show that there might be a small tension with other measurements, such as baryon acoustic oscillations (BAOs) or Supernovae Ia \cite{DiValentino:2019qzk,Handley:2019tkm}. This discordance could be due to unknown systematic effects that may propagate when combinations of different datasets are attempted \cite{Raveri:2018wln}, or to the assumption of a flat fiducial cosmology in the BAO measurements \cite{ODwyer:2019rvi}. In fact, relaxing the flat assumption, BAO data alone seem to favor a closed universe at the $2\sigma$ confidence level, even when they are combined with CMB data \cite{Glanville:2022xes}.

The effect of nonzero spatial curvature on observations is twofold. On the one hand, it affects the background evolution and the CMB power spectrum after inflation, through the radiation, matter and dark energy epochs. In particular, the presence of spatial curvature modifies the transfer function governing the evolution of linear perturbations from early epochs until today \cite{Lewis:1999bs,Masso:2006gv}. The second consequence is related to the corrections to the primordial spectrum of tensor and scalar perturbations that are produced in the inflationary phase. Specifically, possible detection of nonzero curvature would significantly affect the duration of inflation \cite{Lasenby:2003ur}, thus providing an explanation to some unknown features such as the reduced amplitudes on large scales of the CMB spectrum \cite{Efstathiou:2003hk}.

A smoking gun signature for inflation is provided by the generation of a stochastic background of gravitational waves (GWs). In fact, primordial GWs are a key prediction of inflationary models and are not expected in cosmological scenarios that do not consider an inflation mechanism occurring in the early Universe. In simple slow-roll models, primordial GWs are tensor metric fluctuations with a nearly scale-invariant power spectrum at superhorizon scales \cite{Guzzetti:2016mkm}. Besides being a way to probe inflation, the observational signatures of primordial GWs allow us to distinguish among different inflationary models predicting different values of the amplitude of the GW signal.
Also, the importance of primordial GW lies in the fact that the energy scales of the inflationary GW background are much higher than those typically achievable by particle accelerators. Hence, measuring the value of the tensor-to-scalar ratio would probe the energy regime of physics beyond the standard model.

Therefore, the observation of primordial GWs represents a major challenge for cosmology in the next years, as testified by a number of future CMB experiments recently proposed with the aim of detecting the $B$-mode polarization pattern \cite{COrE,PRISM:2013fvg,LiteBIRD:2022cnt}. Furthermore, evidence for primordial GWs could be inferred from tensor modes \cite{Dodelson:2003bv,Masui:2010cz}, or modifications of the cosmic  expansion history \cite{Smith:2006nka}, which may be directly detected through future GW experiments, such as Einstein Telescope (ET) \cite{Maggiore:2019uih,Branchesi:2023mws}, the Laser Interferometer Space Antenna (LISA) \cite{LISA:2017pwj} and the Deci-hertz Interferometer Gravitational Wave Observatory (DECIGO) \cite{Kawamura:2020pcg}. These are expected to play a central role in investigating the nature of gravity, dark energy, and many other fundamental  questions \cite{Cai:2017aea,Belgacem:2017ihm,DAgostino:2019hvh,Bonilla:2019mbm,Baker:2020apq,Tasinato:2021wol,Ezquiaga:2021ler,Califano:2022cmo,Badger:2021enh,DAgostino:2022tdk,Califano:2022syd,Pieroni:2022bbh}. 
Additionally, evidence of GW background could be inferred by observing the GW influence on the arrival time of Pulsar signals on Earth. Nowadays, there are different PTA groups operating worldwide, including the European Pulsar Timing Array (EPTA) \cite{Babak:2015lua,Desvignes:2016yex}, the North American Nanohertz Observatory for Gravitational Waves (NANOGrav) \cite{NANOGRAV:2018hou}, and the Parkes Pulsar Timing Array (PPTA) \cite{PPTA}. These individual groups are also the constituents of the International Pulsar Timing Array (IPTA) collaboration \cite{IPTA}.

The aim of the present paper is to investigate the role of spatial curvature in the propagation of  GWs in the primordial Universe. We are specifically interested in describing the evolution of tensor perturbations at early times and studying the possible implications of nonflat cosmological scenarios on the power spectrum of GWs generated during inflation. This allows us to analyze the energy density and frequency of GWs that can be directly compared with the typical sensitivities of the next-generation GW detectors. 

This work is structured as follows. In Sec.~\ref{sec:background}, we briefly overview the cosmological background evolution of FLRW universes with nonvanishing spatial curvature. In particular, we show the solutions to the Friedmann equations for radiation, matter and de Sitter eras. In Sec.~\ref{sec:perturbations}, we discuss linear perturbations on the FLRW metric and obtain analytical solutions of the tensor wave equation for different spatial geometries.
In Sec.~\ref{sec:power spectrum}, we analyze the effects of spatial curvature on the primordial power spectrum of GWs generated during the inflation epoch. We thus investigate the evolution of transfer functions in the radiation and matter epochs, discussing the matching conditions between the two stages.
Furthermore, in Sec.~\ref{sec:effective}, we study the impact of the effective relativistic degrees of freedom on the spectral density and frequency of GWs. 
In Sec.~\ref{sec:observations}, we discuss the observational consequences of our results in view of potential detections by present and future experiments.
Finally, in Sec.~\ref{sec:conclusions}, we summarize our findings and outline the future perspectives of this work.

Throughout the paper, we set units $c=\hbar=8\pi G=1$.

\section{Background FLRW cosmology}
\label{sec:background}

The standard cosmological model is based on the Einstein-Hilbert gravitational action with the cosmological constant contribution:
\begin{equation}\label{E-H action}
    S=\int d^4x\sqrt{-g}\left[\dfrac{1}{2}\left(R-2\Lambda\right)+\mathcal{L}_m\right].
\end{equation}
Here, $g$ is the determinant of the metric tensor $g_{\mu\nu}$, $R$ is the Ricci scalar and $\mathcal{L}_m$ is the Lagrangian density of matter fields. By varying the above action with respect to $g_{\mu\nu}$, one obtains the Einstein field equations
\begin{equation}\label{eq: Einstein Field Equation}
    R_{\mu\nu}-\dfrac{1}{2}g_{\mu\nu}R+\Lambda g_{\mu\nu}= T_{\mu\nu}\,,
\end{equation}
where $R_{\mu\nu}$ is the Ricci tensor and $T_{\mu\nu}$ is the matter energy-momentum tensor:
\begin{equation}
 T_{\mu\nu}= \frac{-2}{\sqrt{-g}}\frac{\delta(\sqrt{-g}\mathcal{L}_m)}{\delta g^{\mu\nu}}\,.
 \end{equation}
Assuming the Universe to be filled with a perfect fluid of energy density $\rho$ and pressure $p$, one has
\begin{equation} \label{eq: perfetc fluid SE tensor}
    T_{\mu\nu}= \left(\rho+p\right)u_{\mu}u_{\nu}+p g_{\mu\nu}\,,
\end{equation}
where $u_{\mu}$ is the four-velocity of the cosmic fluid. 

Under the cosmological principle, spacetime is described by the homogeneous and isotropic FLRW metric:
\begin{equation}\label{eq:metric}
       ds^2\equiv g_{\mu\nu}dx^\mu dx^\nu=a(\tau)^2\left[-d\tau^2+\gamma_{ij}dx^idx^j\right],
\end{equation}
where $\gamma_{ij}$ is the maximally symmetric metric of spatial hypersurfaces, such that
\begin{equation}
   \gamma_{ij}dx^idx^j=\dfrac{dr^2}{1-K r^2}+r^2d\Omega^2\,,
\end{equation}
being $d\Omega^2\equiv d\theta^2+\sin^2\theta\, d\phi^2$ the infinitesimal solid angle,
and $a$ the dimensionless scale factor as a function of the conformal time\footnote{The conformal time is defined as $d\tau\equiv dt/a$, being $t$ the cosmic (physical) time. In our treatment, we take the scale factor normalized at the present time, namely $a(\tau)\equiv a/a_0$, with $a_0$=1.}. Here, we consider the parameter $K$ to have units of length$^{-2}$, with its
sign identifying the curvature of the three-dimensional space: flat $(K=0)$, closed $(K>0)$ and open $(K<0)$ universes. 

Solving Eq.~\eqref{eq: Einstein Field Equation} for the metric \eqref{eq:metric} provides us with the Friedmann equations describing the background evolution of the Universe: 
\begin{align}\label{eq: friedmann 1}
     \mathcal{H}^2&=\dfrac{1}{3}\left(\rho +\Lambda\right)a^2-K,\\
   \mathcal{H}'+\mathcal{H}^2&=\dfrac{1}{6}\left(\rho-3p\right)a^2+\dfrac{2 \Lambda }{3}a^2-K,\label{eq: friedmann 2}
\end{align}
where $\mathcal{H}\equiv a'/a$ is the conformal Hubble parameter, and the symbol $^\prime$ denotes the derivative with respect to $\tau$.

For each species of the cosmic fluid, $i$, we can assume a linear barotropic equation of state $p_i=w_i\rho_i$, so that the conservation of the energy-momentum tensor, $\nabla^\nu T_{\mu\nu}=0$, provides us with the continuity equation
\begin{equation}
\rho_i^{\prime}+3\mathcal{H}(1+w_i)\rho_i=0\,,
\end{equation}
whose solution is
\begin{equation}\label{eq: continuity eq rho}
    \rho_i=\rho_{0,i}a^{-3(1+w_i)},
\end{equation}
being $\rho_{0,i}$ the present-day energy density of each species.

The Friedmann equations can be thus solved by means of Eq.~\eqref{eq: continuity eq rho}, together with the initial condition $a(0)=0$. 
Specifically, in the radiation epoch, setting $w=1/3$ and $\Lambda=0$, one finds
\begin{numcases}{a_\text{rad}(\tau)\propto}
  \sinh\left(\tau\sqrt{|K|}\right)\,, \hspace{0.3cm} &$K<0$\,,\label{eq: radOpen}\\ 
       \tau\,, \hspace{0.3cm} &$K=0$\,,\label{eq: radFlat}\\
\sin\left(\tau\sqrt{K}\right)\,, \hspace{0.3cm} &$K>0$\,. \label{eq: radClosed}
\end{numcases}

In the epoch dominated by nonrelativistic matter, the Friedmann equations for $w=0$ and $\Lambda=0$ are solved for
    \begin{numcases}{ a_\text{mat}(\tau)\propto}
\cosh \left(\tau\sqrt{|K|}\right)-1, \hspace{0.3cm} & $K<0$\,,\label{eq: matOpen}\\
       \tau^2\,, \hspace{0.3cm} & $K=0$\,,\label{eq: matFlat}\\
    1-\cos\left(\tau\sqrt{K}\right), \hspace{0.3cm} & $K>0\,.$\label{eq: matClosed}
    \end{numcases}    

Regarding the inflationary dynamics, we consider a de Sitter universe driven by the cosmological constant $(w=-1)$, with nonvanishing spatial curvature. In this case, the solution to the scale factor is
\begin{numcases} {a_{\inf}(\tau)=}
      -\dfrac{\sqrt{|K|}}{ \mathcal H_{\Lambda}\sinh\left(\tau\sqrt{|K|}\right)}\,, \hspace{0.3cm} &$K<0$\,,\label{eq: openINF}\\  -\dfrac{1}{\mathcal H_{\Lambda}\tau }\,, \hspace{0.3cm} &$K=0$\,,\label{eq: flatINF}\\
      -\dfrac{\sqrt{K}}{\mathcal H_{\Lambda}\sin\left(\tau\sqrt{K}\right)}\,,\hspace{0.3cm} &$K>0$\,, \label{eq: closedINF}
\end{numcases}
where $\mathcal H_{\Lambda}=\sqrt{\Lambda/3}$. It is worth noticing that, in both open and flat universes, the conformal time domain is $\tau=(-\infty,0)$, and the scale factor is defined over the interval $a=(0,\infty)$. In the closed case, instead, the conformal time spans the range $\tau=(-\pi/(2\sqrt{K}),0)$, while the scale factor domain is $a=(\sqrt{K}/\mathcal{H}_\Lambda, \infty$). 

In the next section, we discuss linear perturbations over the FLRW background. In particular, we shall study the influence of spatial curvature on the power spectrum of tensor perturbations generated by inflation.

\section{Tensor Perturbations}
\label{sec:perturbations}

In order to describe the propagation of GWs during inflation, we consider linear perturbations around the spatial part of the line element \eqref{eq:metric}:
\begin{equation}
    ds^2= a(\tau)^2\left[-d\tau^2+(\gamma_{ij}+h_{ij})dx^idx^j\right],
    \label{perturbed metric}
\end{equation}
where $h_{ij}$ is a symmetric 3-tensor satisfying $h_i^i=\mathcal{D}_ih_j^i=0$, with $\mathcal{D}_i$ denoting the covariant derivative compatible with $\gamma_{ij}$\footnote{The indices of the tensor perturbation $h_{ij}$ are lowered and raised through the background metric $\gamma_{ij}$. They are invariant under gauge and conformal transformations.}. For the gravitational theory \eqref{E-H action}, the second-order action for tensor perturbations is given as \cite{Mukhanov:1990me}
\begin{equation}\label{eq: second order action}
    S^{(2)}=\frac{1}{8}\int d^4x \sqrt{\gamma}\, a^2\left(h^{ij\prime}h_{ij}-\mathcal{D}_l h_{ij}\mathcal{D}^l h^{ij}-2Kh^{ij}h_{ij}\right),
\end{equation}
where $\gamma\equiv\det(\gamma_{ij})$. Thus, the equation of motion is obtained by varying action~\eqref{eq: second order action} with respect to $h_{ij}$:
\begin{equation}\label{GW evolution}
    h^{\prime\prime}_{ij}+2\mathcal{H}h^{\prime}_{ij}+2Kh_{ij}=\mathcal{D}^2 h_{ij},
\end{equation}
where $\mathcal{D}^2\equiv\gamma^{ij}\mathcal{D}_i\mathcal{D}_j$. It is worth to stress that the above equation describes the evolution of GWs along all phases of the Universe's history.

For a spatially flat geometry, tensor perturbations can be expanded in eigenfunctions of the flat-space Laplacian operator. Analogously, in a spatially curved universe, it is convenient to expand the $h_{ij}$ perturbations in terms of tensor harmonics $Q^{nlm(s)}_{ij}(\bm{y})$, with $\bm{y}\equiv(r,\,\theta,\,\phi)$, such that \cite{Abbott:1986ct}
\begin{equation}\label{eq: laplacian}
    \mathcal{D}^2Q^{nlm(s)}_{ij}=-(n^2-3K)Q^{nlm(s)}_{ij}\,,
\end{equation}
satisfying the conditions $\gamma^{ij}Q^{nlm(s)}_{ij}=\mathcal{D}^i Q^{nlm(s)}_{ij}=0$. Here, $n$ represents the wavenumber generalized to a spatially curved geometry\footnote{More generally, the curved-space wavenumber can be written in terms of the rank of the perturbation type, $\beta$, as $n=\sqrt{k^2+(\beta+1)K}$, where $\beta=0,\, 1,\, 2$ for scalars, vectors and tensors, respectively.}:
\begin{equation}
    n=\sqrt{k^2+3K}\,.
    \label{eq:generalized wavenumber}
\end{equation}
The latter reduces to the flat eigenmode $k$ as $K\rightarrow 0$.

It can be shown that the spectrum of the tensor harmonics is complete for $n\geq 0$, with $l\geq 2$ and $-l\leq m \leq l$, in the case $K\leq 0$, whereas for $n/\sqrt{K}=3,4,5,\hdots$\,, with $2\leq l\leq n-1$ and $-l\leq m \leq l$, in the case $K>0$ (see Refs.~\cite{Abbott:1986ct,Hu:1997mn,Akama:2018cqv} for the details).
Moreover, the index $s$ accounts for the harmonics parity, and the normalization condition for the tensor harmonics reads
\begin{equation}
    \int \dfrac{r^2 dr\, d\Omega}{\left(1+\frac{K}{4}r^2\right)^3}Q_{ij}^{nlm}(\bm{y})^* Q_{ij}^{n'l'm'}(\bm{y})=\dfrac{\pi}{2}\delta(n,n')\delta_{ll'}\delta_{mm'}\,,
\end{equation}
where the function $\delta(n,n')$ is defined with respect to the measure $\mu(n)$:
\begin{equation}
    \int d\mu\, f(n)\, \delta(n,n')=f(n')\,,
\end{equation}
being $f(n)$ a generic function of the index $n$.
Both the function $\delta(n,n')$ and the measure $\mu(n)$ are defined depending on the spatial curvature:
\begin{equation}
\delta(n,n')=
    \begin{cases}
        \delta(n-n')\,, &\quad K\leq 0\,, \\
        \delta_{nn'}\,, &\quad K>0\,,
   \end{cases}
   \end{equation}
and
\begin{equation}
    \int d\mu= 
    \begin{cases}
        \displaystyle \int_0^\infty dn\,, &\quad K\leq 0\,, \vspace{0.2cm}   \\
        \displaystyle \sum_{n/\sqrt{K}=3}^\infty\,, &\quad K>0\,.
    \end{cases}
    \label{eq:measure}
\end{equation}

The tensor perturbations can be thus expanded as
\begin{equation}
     h_{ij}(\tau,\bm{y})=\sum_s\sum_{nlm}h_{nlm}^{(s)}(\tau)\, Q_{ij}^{nlm(s)}(\bm{y}) \,,
     \label{eq:tensor decomposition}
\end{equation}
where the sum over $n$ should be understood as the right-hand sides of Eq.~\eqref{eq:measure}.
It is worth noticing that $h_{nlm}^{(s)}(\tau)$ encodes all the dynamical information of GW evolution.
Taking into account Eqs.~\eqref{eq: laplacian} and \eqref{eq:tensor decomposition}, 
we can write Eq.~\eqref{GW evolution} in the form
\begin{equation}\label{eq: curved GW eq}
     h_{nlm}^{(s)\prime\prime}+2\mathcal{H}h_{nlm}^{(s)\prime}+\left(n^2-K\right)h_{nlm}^{(s)}=0\,.
\end{equation}
Then, it is convenient to define the quantity
\begin{equation}
    \sigma_{nlm}^{(s)}(\tau)\equiv a(\tau)\,h_{nlm}^{(s)}(\tau)\,,
    \label{eq: rescaling inflation}
\end{equation}
which can be promoted to the quantum operator
\begin{equation}
\hat{\sigma}_{nlm}^{(s)}(\tau)=\sigma_{nlm}^{(s)}(\tau)\,\hat{\mathfrak a}_{nlm}^{(s)}+\sigma_{nlm}^{(s)*}(\tau)\,\hat{\mathfrak a}_{nlm}^{(s)\dagger}\,,
\end{equation}
where $\hat{\mathfrak a}_{nlm}^{(s)\dagger}$ and $\hat{\mathfrak a}_{nlm}^{(s)}$ are the creation and annihilation operators, respectively, satisfying the following commutation rules:
\begin{align}
    &[\hat{\mathfrak a}_{nlm}^{(s)},\, \hat{\mathfrak{a}}_{n'l'm'}^{(s')\dagger}]=\delta(n,n')\delta_{ss'}\delta_{ll'}\delta_{mm'}\,,
    \\
    &[\hat{\mathfrak a}_{nlm}^{(s)},\, \hat{\mathfrak{a}}_{n'l'm'}^{(s')}]=[\hat{\mathfrak a}_{nlm}^{(s)\dagger},\, \hat{\mathfrak{a}}_{n'l'm'}^{(s')\dagger}]=0\,.
\end{align}
The vacuum state in the Hilbert space is defined through the standard condition 
\begin{equation}
\hat{\mathfrak{a}}^{(s)}_{nlm}\ket{0}=0\,,
    \label{eq:vacuum state}
\end{equation}
and the quantum excited states are produced by multiple applications of the operator $\hat{\mathfrak a}_{nlm}^{(s)\dagger}$. 
Hence, using Eq.~\eqref{GW evolution} and Eq.~\eqref{eq:generalized wavenumber}, we obtain 
\begin{equation}
\sigma_{nlm}^{(s)\prime\prime}+\left(n^2-K-\dfrac{a''}{a}\right)\sigma_{nlm}^{(s)}=0\,,
\label{eq:master}
\end{equation}
where the eigenmodes $\sigma_{nlm}^{(s)}(\tau)$ satisfy the normalization condition 
\begin{equation}
    \sigma_{nlm}^{(s)} \sigma_{nlm}^{(s)*\prime}-\sigma_{nlm}^{(s)\prime}\sigma_{nlm}^{(s)*}=i\,.
    \label{eq:normalization}
\end{equation}

In the next paragraphs, we shall search for the solutions of Eq.~\eqref{eq:master} for different spatial curvatures.
As the dynamics of the GW equation is influenced only by the generalized wavenumber $n$, in what follows we will drop the superscript $(s)$ as well as the indices $l$ and $m$ to simplify the notation.

\subsection{Flat universe}

Before addressing the open and closed universe cases, we first review the evolution of tensor perturbations in a flat de Sitter universe \cite{Watanabe:2006qe,Guzzetti:2016mkm,Baumann:2018muz}. In this case, 
inserting the solution~\eqref{eq: flatINF} into Eq.~\eqref{eq:master} provides
\begin{equation}\label{eq: flat inflationary rescaled eq}
    \sigma_n''(\tau)+\left(n^2-\frac{2}{\tau^2}\right)\sigma_n(\tau)=0\,,
\end{equation}
which admits the general solution
\begin{equation}
    \label{eq: solution flat inflation}
    \sigma_n(\tau)=c_1e^{in\tau}\left(1+\dfrac{i}{n \tau}\right)+c_2e^{-in\tau}\left(1-\dfrac{i}{n \tau}\right).
\end{equation}
Here, $c_1$ and $c_2$ are complex integration constants to be fixed through suitable initial conditions. In particular, imposing Eq.~\eqref{eq:normalization} leads to
\begin{equation}
    |c_2|^2-|c_1|^2=\frac{1}{2n}\,.
    \label{eq:constraint flat}
\end{equation}
This, however, is not sufficient to uniquely determine the eigenmode $\sigma_n(\tau)$, as a variation of the latter would imply a variation of $\hat{\mathfrak a}_n$ such that $\hat\sigma_n(\tau)$ remains unchanged. In turn, each of these solutions corresponds to a different vacuum. Nevertheless, requiring the vacuum to coincide with the lowest-energy eigenstate of the Hamiltonian at some particular time $\tau=\tilde{\tau}$, it is possible to select a specific solution of $\sigma_n(\tau)$. 
A standard choice is the Bunch-Davies vacuum \cite{Bunch:1978yq,Birrell:1982ix}, for which the ground state of the Hamiltonian is defined in the infinite past, i.e., $\tilde{\tau}\rightarrow -\infty$. In this limit, one finds the asymptotic behavior $\sigma_n\sim e^{-in\tau}$. Comparing the latter to Eq.~\eqref{eq: solution flat inflation}, and making use of Eq.~\eqref{eq:constraint flat}, we obtain the conditions
\begin{equation}
 c_1=0\,, \quad   |c_2|^2=\frac{1}{2n}\,,
\end{equation}
which finally leads to 
\begin{equation}
    \sigma_n(\tau)=\frac{e^{-in\tau}}{\sqrt{2n}}\left(1-\dfrac{i}{n \tau}\right).
    \label{eq:sol_flat}
\end{equation}

\subsection{Open universes}

In the case of open universes ($K<0$), the inflationary behavior of tensor perturbations is given by inserting the background solution~\eqref{eq: openINF} into Eq.~\eqref{eq:master}, so to obtain
\begin{equation}
\label{eq: rescaled inflationary eq flat}
\sigma_n''(\tau)+\left[n^2-\dfrac{2|K|}{\sinh^2{\left(\tau\sqrt{|K|}\right)}}\right]\sigma_n(\tau)=0\,.
\end{equation}
The general solution to the latter is
\begin{equation}
    \label{eq: solution open}
    \begin{split}
   \sigma_n(\tau)=&\ c_1e^{i n\tau}\left[n+i\sqrt{|K|} \coth\left(\tau\sqrt{|K|}\right)\right]\\
   & +c_2e^{-i n\tau}\left[n-i\sqrt{|K|} \coth\left(\tau\sqrt{|K|}\right)\right], 
    \end{split}
\end{equation}
where the constants of integration $c_1$ and $c_2$ may be fixed by imposing the Bunch-Davies vacuum state in the infinite past limit, $\tilde{\tau}\rightarrow-\infty$, along with normalization \eqref{eq:normalization}. These imply 
\begin{equation}
   c_1=0\,, \quad |c_2|^2=\dfrac{1}{2n \left(n^2+|K|\right)}\,,
\end{equation}
which provide
\begin{equation}
   \sigma_n(\tau)=\dfrac{e^{-i n\tau}}{\sqrt{2n \left(n^2+|K|\right)}}\left[n-i\sqrt{|K|} \coth\left(\tau\sqrt{|K|}\right)\right].
   \label{eq:sol_open}
\end{equation}
As a consistency check, one can easily verify that the above expression recovers the flat case solution \eqref{eq:sol_flat} in the limit $K\rightarrow 0$.

\subsection{Closed universes}

A simple way to obtain solutions of the tensor perturbation equation for closed universes $(K>0)$ is through the transformations $\tau \rightarrow i \tau$ and $ n\rightarrow i n$ to be applied to the equations governing the dynamics of open universes.
Indeed, using this mapping into Eq.~\eqref{eq: rescaled inflationary eq flat}, we find
\begin{equation}
    \sigma_n''(\tau)+\left[n^2-\dfrac{2K}{\sin^2\left(\tau\sqrt{K}\right)}\right] \sigma_n(\tau)=0\,,
\end{equation}
which is consistent with the result one would obtain from Eqs.~\eqref{eq: matClosed} and \eqref{eq:master}. Then, applying the same transformations to Eq.~\eqref{eq: solution open} immediately provides us with the general solution
\begin{equation}
\label{eq:gen_sol_closed}
\begin{split}
       \sigma_n(\tau)=&\ c_1e^{i n\tau}\left[n+i\sqrt{K} \cot\left(\tau\sqrt{K}\right)\right]\\
       &+c_2e^{-i n\tau}\left[n-i\sqrt{K} \cot\left(\tau\sqrt{K}\right)\right].
\end{split}
\end{equation}
However, differently from the flat and open cases, the Bunch-Davies vacuum condition is here realized in the limit $\tilde{\tau}\rightarrow-\frac{\pi}{2\sqrt{K}}$, which represents the remote past of closed inflationary models. Hence, the integration constants appearing in Eq.~\eqref{eq:gen_sol_closed} may be fixed to 
\begin{equation}
    c_1=0\,, \quad |c_2|^2=\dfrac{1}{2n(n^2-K)}\,,
\end{equation}
which leads to the final expression
\begin{equation}
    \sigma_n(\tau)=\dfrac{e^{-i n\tau}}{\sqrt{2n(n^2-K)}}\left[n-i\sqrt{K} \cot\left(\tau\sqrt{K}\right)\right] .
    \label{eq:sol_closed}
\end{equation}
We notice that the above expression reduces to Eq.~\eqref{eq:sol_flat} in the limit $K\rightarrow 0$.

\section{Power Spectrum of primordial Gravitational Waves}
\label{sec:power spectrum}

In this section, we analyze the physical consequences of the results obtained for different spatial FLRW geometries. In particular, we are interested in studying the primordial power spectrum of GWs and the evolution of transfer functions \cite{Boyle:2005se,Watanabe:2006qe,Guzzetti:2016mkm}. 

The comoving wavenumber $n$ is associated with its corresponding physical wavelength, $\lambda$, through $\lambda\sim a/n$.
The latter can be further related to the conformal Hubble radius, $a\mathcal{H}^{-1}$, defining the size of causally connected regions at a given epoch. 
The behavior of perturbations can be therefore characterized by two particular regimes: 
the subhorizon regime ($\lambda\ll a \mathcal{H}^{-1}$),  describing perturbations inside the horizon that are damped by the expansion of the Universe; the superhorizon regime ($\lambda\gg a \mathcal{H}^{-1}$), where perturbations have exited the horizon and get stretched by the inflation mechanism.
In terms of the comoving wavenumber, the subhorizon limit can be expressed by the condition $n\gg \mathcal{H}$, whereas the super-horizon limit as $n\ll \mathcal{H}$, with the horizon crossing occurring as $n=\mathcal{H}$. In this context, the presence of nonvanishing spatial curvature plays an important role, as it can anticipate or delay the horizon crossing of a given mode with respect to the flat case.

As previously discussed, Eq.~\eqref{GW evolution} describes the evolution of tensor modes in the absence of sources. Labeling as $h_n^\text{inf}$ the amplitude of primordial GWs that left the horizon during the inflationary phase, 
the general solution of tensor perturbations, valid at any time, may be written in the form \cite{Watanabe:2006qe}
\begin{equation}\label{eq: factorized solution}
    h_n(\tau)\equiv h_{n}^\text{inf}\mathscr{T}(\tau,n)\,,
\end{equation}
where $\mathscr{T}(\tau,n)$ is the transfer function characterizing the GW modes evolution after entering the horizon in the post-inflationary universe. The transfer function is normalized such that $\mathscr{T}(\tau,n)\rightarrow 1$ as $n\rightarrow 0$,  and its evolution is described by
\begin{equation}
    \mathscr{T}''(\tau,n)+2\mathcal{H}\mathscr{T}'(\tau,n)+\left(n^2-K\right)\mathscr{T}(\tau,n)=0\,, \label{eq: curved GW eq TF}
\end{equation}
which is subjected to the boundary conditions \cite{Bernal:2019lpc}
\begin{equation}
    \mathscr{T}(0,n)=1\,,   \quad  \mathscr{T}'(0,n)=0\,.
    \label{eq:initial_TF}
\end{equation}
Two different epochs are taken into account by Eq.~\eqref{eq: curved GW eq TF}: the radiation era, for $\tau<\tau_\text{eq}$ ($n>n_\text{eq}$), and the matter era, for $\tau>\tau_\text{eq}$ ($n<n_\text{eq}$), with $\tau_{\rm eq}$ being the equivalence time between the two epochs \cite{Kite:2021yoe}.

Then, one may define the tensor power spectrum of primordial GWs 
as
\begin{equation}
    \mathcal{P}_T(n)\equiv \dfrac{n^3}{\pi^2}\Big[|h^{\text{inf}}_{n}|^2\Big]_{\tau=\tau_\star}\,, 
    \label{eq: PPS2}
\end{equation}
where $\tau_\star$ denotes the time of the horizon crossing, i.e., when $n=\mathcal{H}$.
Furthermore, if we consider the time-time component of the GW energy-momentum tensor,
\begin{equation}
    \rho_h(\tau)=\dfrac{\langle h_{ij}'(\tau,\bm{y})h^{ij\prime}(\tau,\bm{y})\rangle}{4a(\tau)^2}\,,
\end{equation}
we can express the relative spectral energy density of GWs as
\begin{equation}
\Omega_\text{GW}(\tau,n)\equiv \dfrac{\tilde\rho_h(\tau,n)}{\rho_\text{cr}(\tau)}=
\dfrac{\mathcal{P}_T(n)}{12 \mathcal{H}(\tau)^2}\left|\mathscr{T}^{\prime}(\tau,n)\right|^2\,,
\label{eq: spectral energy density}
\end{equation}
where $\tilde{\rho}_h(\tau,n)=\frac{d\rho_h}{d\ln n}$, and $\rho_\text{cr}(\tau)\equiv 3\mathcal{H}(\tau)^2 a(\tau)^{-2}$ is the critical density of the Universe. We refer the reader to Ref.~\cite{Watanabe:2006qe} for the details.

In the following, we shall distinguish among different spatial curvatures to derive the GW power spectrum and analyze the evolution of the  transfer functions in the radiation and matter eras.

\subsection{Flat universe}

Let us start to calculate the primordial power spectrum for a flat universe. Inverting Eq.~\eqref{eq: friedmann 1} for $n=\mathcal{H}$, we find the scale factor at the horizon crossing:
\begin{equation}
    a_\star=\dfrac{n}{\mathcal H_\Lambda}\,.
\end{equation}
This can be used with Eq.~\eqref{eq: flatINF} to obtain the conformal time at the horizon crossing:
\begin{equation}
    \tau_\star=-\dfrac{1}{n}\,.
\end{equation}
Hence, combining the above expressions with Eqs.~\eqref{eq: rescaling inflation} and \eqref{eq:sol_flat}, the power spectrum \eqref{eq: PPS2} reads
\begin{equation}\label{eq: PPS flat}
  \mathcal{P}_T= \left(\dfrac{\mathcal H_{\Lambda}}{\pi}\right)^2,
\end{equation}
which is clearly a scale-invariant quantity.

The evolution of transfer functions after inflation in a flat universe is obtained by solving Eq.~\eqref{eq: curved GW eq TF} with $K=0$. 
In particular, we consider the background solutions given by Eqs.~\eqref{eq: radFlat} and \eqref{eq: matFlat}, and introduce the new variable $u\equiv n\tau$. Hence, Eq.~\eqref{eq: curved GW eq TF} reads
\begin{equation}
    \mathscr{T}''(u)+\dfrac{2\alpha}{u}\mathscr{T}'(u)+\mathscr{T}(u)=0\,,
    \label{eq:transf flat}
\end{equation}
where $\alpha=1,2$ for radiation and matter epochs, respectively, and the prime denotes the derivative with respect to $u$.
The initial conditions associated to Eq.~\eqref{eq:transf flat} are
 \begin{equation}
 \label{eq: TF flat initial conditions}
    \mathscr{T}(0)= 1\,, \quad \mathscr{T}'(0)=0\,.
    \end{equation}
The general solution in the radiation era is 
\begin{equation}
    \mathscr{T}_\text{rad}(u)=A\frac{e^{-i u}}{u}+B\frac{e^{i u}}{u}\,,
    \label{eq:gen_TF_flat_RD}
\end{equation}
where $A$ and $B$ are complex coefficients to be determined.
The transfer functions are real quantities, so we require the condition $\mathscr{T}(u)=\mathscr{T}(u)^*$. 
As Eq.~\eqref{eq:gen_TF_flat_RD} may be formally written as $\mathscr{T}=A f+B f^*$, the above condition translates into $(A-B^*)f=(A^*-B)f^*$, which holds true only if $A=B^*$ and $A^*=B$. If we write $A=a_1+ia_2$ and $B=b_1+ib_2$, then $b_1=a_1$ and $b_2=-a_2$. Thus, after a convenient redefinition of the constants, we obtain
\begin{equation}
    \mathscr{T}_\text{rad}(u)=c_1\frac{\sin u}{u}+c_2\frac{\cos u}{u}\,,
\end{equation}
where $c_1$ and $c_2$ are now real constants, which can be found by requiring the initial conditions \eqref{eq: TF flat initial conditions}. In particular, we find $c_1=1$ and $c_2=0$, so we finally have
\begin{equation} 
    \mathscr{T}_\text{rad}(u)=\dfrac{\sin u}{u}\,. 
    \label{eq: Flat RD TF}
\end{equation}

In the matter era, the general solution of Eq.~\eqref{eq:transf flat} can be written as
\begin{equation} 
    \mathscr{T}_\text{mat}=c_1\left(\frac{\sin u}{u^3}-\dfrac{\cos u}{u^2}\right)+c_2 \left(\frac{\sin u}{u^2}+\frac{\cos  u}{u^3}\right),
    \label{eq:gen_TF_flat_MD}
\end{equation}
where $c_1$ and $c_2$ are real coefficients.
In this case, the initial conditions \eqref{eq: TF flat initial conditions} are satisfied for $c_1=3$ and $c_2=0$, leading to
\begin{equation}
     \mathscr{T}_\text{mat}(u)=\frac{3}{u^3}\left(\sin u-u \cos u\right)\,.
     \label{eq: Flat MD TF}
\end{equation}

In order to find a smooth matching between the two epochs, we need to consider the propagation of radiation modes into the matter era. This can be achieved by taking into account the general solution in the intermediate regime $(\tau>\tau_\text{eq},n>n_\text{eq})$:
\begin{equation}
     \mathscr{T}_\text{int}=\dfrac{u_\text{eq}}{u}\left[A_\text{eq}\left(\frac{\sin u}{u^2}-\dfrac{\cos u}{u}\right)+B_\text{eq} \left(\frac{\sin u}{u}+\frac{\cos  u}{u^2}\right)\right].
     \label{eq: TF inter flat}
\end{equation}
Here, the constants $A_\text{eq}$ and $B_\text{eq}$ can be determined by requiring the matching between Eqs.~\eqref{eq: Flat RD TF} and \eqref{eq: TF inter flat}, together with their respective first derivatives, evaluated at the time of equivalence.
In so doing, one finds
\begin{align}\label{eq: A consst}
    A_\text{eq}&=1-\frac{1}{u_\text{eq}^2}+\frac{\sin (2 u_\text{eq})}{2 u_\text{eq}}+\frac{\cos (2 u_\text{eq})}{u_\text{eq}^2}\,, \\
    B_\text{eq}&=\frac{3}{2 u_\text{eq}}+\frac{\sin (2 u_\text{eq})}{u_\text{eq}^2}-\frac{\cos (2 u_\text{eq})}{2 u_\text{eq}}\,.
    \label{eq: B consst}
\end{align}
It is worth noting that the solutions \eqref{eq: Flat RD TF}, \eqref{eq: Flat MD TF} and \eqref{eq: TF inter flat} may be  also written in terms of spherical Bessel functions of the first kind \cite{Watanabe:2006qe}.

\subsection{Open universes}

Considering the case of open universes, the scale factor at the horizon crossing is given from Eq.~\eqref{eq: friedmann 1} as
\begin{equation}
    a_\star=\frac{\sqrt{n^2-|K|}}{\mathcal H_\Lambda}\,.
    \label{eq:scale factor hc nonflat}
\end{equation}
To obtain the expression of the primordial power spectrum for open models, we invert Eq.~\eqref{eq: openINF} with the help of Eq.~\eqref{eq:scale factor hc nonflat}. In so doing, one finds the conformal time at the horizon crossing:
\begin{equation}
    \tau_\star=-\dfrac{1}{\sqrt{| K |}}\text{arctanh} \left(\frac{\sqrt{| K | }}{n}\right).
\end{equation}
Plugging the above expressions into Eqs.~\eqref{eq: rescaling inflation} and \eqref{eq:sol_open}, from Eq.~\eqref{eq: PPS2} we obtain
\begin{equation}
\mathcal{P}_T(n)=\left(\dfrac{\mathcal H_\Lambda}{\pi}\right)^2\dfrac{n^4}{n^4-K^2}\,.
\label{eq:PS open}
\end{equation}
The latter suggests that, in a universe with spatial hyperbolic geometry, the primordial power spectrum of tensor perturbations produced during the inflation results to be enhanced with respect to the spatially flat case, especially for the smallest modes (namely, for the longest physical wavelengths). As soon as $n\gg \sqrt{|K|}$, Eq.~\eqref{eq:PS open} smoothly approaches the scale-invariant power spectrum \eqref{eq: PPS flat}. 

In order to derive the transfer functions in the post-inflationary universe, it is convenient to perform the rescaling 
\begin{equation}
    \mathcal{S}(\tau,n)\equiv a(\tau) \mathscr{T}(\tau,n)\,.
    \label{eq:resc_TF}
\end{equation}
Then, Eq.~\eqref{eq: curved GW eq TF} becomes
\begin{equation}
    \mathcal{S}''(\tau,n)+\left[n^2-K-\dfrac{a''(\tau)}{a(\tau)}\right]\mathcal{S}(\tau,n)=0\,,
     \label{eq:TF rescaled}
\end{equation}
obeying the initial conditions
\begin{equation}
    \mathcal{S}(0,n)=0\,, \quad \mathcal{S}'(0,n)=1\,.
    \label{eq:initial TF rescaled}
\end{equation}

In the radiation era, by inserting the scale factor \eqref{eq: radOpen} into Eq.~\eqref{eq:TF rescaled}, we  find
\begin{equation}
    \mathcal{S}_\text{rad}''(\tau,n)+n^2\mathcal{S}_\text{rad}(\tau,n)=0\,,
\end{equation}
which can be solved by means of the initial conditions \eqref{eq:initial TF rescaled}, so as to obtain
\begin{equation}
    \mathcal{S}_\text{rad}(\tau,n)=\frac{\sin(n \tau)}{n}\,.
\end{equation}
Therefore, we finally have
\begin{equation}
    \mathscr{T}_\text{rad}(\tau,n)=\frac{\sqrt{| K | } \sin (n \tau)}{n \sinh\left(\tau \sqrt{| K | }\right)}\,.
    \label{eq:TF_RD_open}
\end{equation}
It is straightforward to show that the above equation recovers the flat-case limit \eqref{eq: Flat RD TF} as $K\rightarrow 0$.

In the matter era, the dynamics of tensor perturbations is described by inserting the background solution \eqref{eq: matOpen} into Eq.~\eqref{eq:TF rescaled}, leading to
\begin{equation}
\mathcal{S}_\text{mat}''(\tau,n)+ \left[n^2+\frac{| K | }{1-\cosh \left(\tau \sqrt{| K | }\right)}\right]\mathcal{S}_\text{mat}(\tau,n)=0\,.
\end{equation}
The latter admits the general solution
\begin{align}
    \mathcal{S}_\text{mat}(\tau,n)=&\ A e^{i n \tau} \left[n+i\dfrac{\sqrt{|K|}}{2} \coth \left(\frac{\tau \sqrt{| K | }}{2}\right)\right]   \label{eq:gen_TF_MD} \\
   & +B e^{-i n \tau} \left[n-i\dfrac{\sqrt{|K|}}{2} \coth \left(\frac{\tau \sqrt{| K | }}{2}\right)\right],  \nonumber
\end{align}
where $A$ and $B$ are complex coefficients.
Following the same prescription adopted in the context of the flat case, one can show that the  condition of reality for $\mathcal{S}(\tau,n)$ implies writing Eq.~\eqref{eq:gen_TF_MD} in the form
\begin{align}
    &\mathcal{S}_\text{mat}=c_1 \left[n \sin (n \tau)+\frac{\sqrt{| K | }}{2}  \cos (n \tau) \coth \left(\frac{\tau \sqrt{| K | }}{2}\right)\right] \nonumber \\
    &+c_2 \left[n \cos (n \tau)-\frac{\sqrt{| K | }}{2}  \sin (n \tau) \coth \left(\frac{\tau \sqrt{| K | }}{2}\right)\right]
    \label{eq:S_MD_open}
\end{align}
where $c_1$ and $c_2$ are  real constants. 

Therefore, in view of Eq.~\eqref{eq:resc_TF}, we obtain
\begin{align}
    &\mathscr{T}_\text{mat}=\frac{c_1 \left[2 n | K |  \sin (n \tau)+| K | ^{3/2} \cos (n \tau) \coth \left(\frac{\tau \sqrt{| K | }}{2}\right)\right]}{2 n \left[\cosh \left(\tau \sqrt{| K | }\right)-1\right]} \nonumber \\
    &+\frac{c_2 \left[2 n | K |  \cos (n \tau)-| K | ^{3/2} \sin (n \tau) \coth \left(\frac{\tau \sqrt{| K | }}{2}\right)\right]}{2 n \left[\cosh \left(\tau \sqrt{| K | }\right)-1\right]}.
    \label{eq:gen_sol_IR_open}
\end{align}
To determine $c_1$ and $c_2$, we require the initial conditions \eqref{eq:initial_TF} to be satisfied. For this purpose, we must impose 
\begin{equation}
    c_1=0\,, \quad c_2=-\dfrac{6}{4n^2+|K|}\,,
\end{equation}
so that the final expression  reads
\begin{equation}
    \mathscr{T}_\text{mat}=\frac{3 | K | ^{3/2} \sin (n \tau) \coth \left(\frac{\tau \sqrt{| K | }}{2}\right)-6 n | K |  \cos (n \tau)}{n \left(4 n^2+|K|\right) \left[\cosh \left(\tau \sqrt{| K | }\right)-1\right]}\,.
    \label{eq:TF_MD_open}
\end{equation}
One can show that the latter reduces to Eq.~\eqref{eq: Flat MD TF} in the limit $K\rightarrow 0$.

Then, to study the behavior of the radiation modes approaching the matter era, we consider the general solution in the intermediate regime:
\begin{equation}
    \mathscr{T}_\text{int}=A_\text{eq}\frac{P_1(\tau)}{Q(\tau)}+B_\text{eq}\frac{P_2(\tau)}{Q(\tau)}\,,
    \label{eq:TF_IR_open}
\end{equation}
where 
\begin{subequations}
\begin{align}
P_1(\tau)&\equiv 2 n | K |  \sin (n\tau)+| K | ^{3/2} \cos (n\tau) \coth \left(\tau\sqrt{|K|}/2\right), \\
P_2(\tau)&\equiv 2 n | K |  \cos (n\tau)-| K | ^{3/2} \sin (n\tau) \coth \left(\tau\sqrt{|K|}/2\right),\\
Q(\tau)&\equiv \cosh\left(\tau \sqrt{|K|}\right)-1\,.
\end{align}
\end{subequations}
We thus require a smooth matching between the radiation and matter solutions by equating Eq.~\eqref{eq:TF_RD_open} to Eq.~\eqref{eq:TF_IR_open}, and their respective derivatives, evaluated at the equivalence epoch. This allows us to find
\begin{equation}
    A_\text{eq}=  \frac{4 n^2 \tanh \left(\frac{\tau_\text{eq} \sqrt{| K | }}{2}\right)-\frac{4 | K |  \sin ^2(n \tau_\text{eq})}{\sinh \left(\tau \sqrt{| K | }\right)}+\frac{n \sqrt{| K | } \sin (2 n \tau_\text{eq})}{\cosh ^2\left(\tau_\text{eq} \sqrt{| K |}/2\right)}}{2 n^2 \sqrt{| K | } \left(4 n^2+| K | \right)},
\end{equation}
\begin{equation}
     B_\text{eq}=\frac{\frac{n \left[\cos (2 n \tau_\text{eq})-\cosh \left(\tau_\text{eq}\sqrt{|K|}\right)-2\right]}{\cosh ^2\left(\tau_\text{eq}\sqrt{|K|}/2\right)}-\frac{2 \sqrt{| K | } \sin (2 n \tau_\text{eq})}{\sinh (\tau_\text{eq}\sqrt{|K|})}}{2 n^2 \left(4 n^2+| K |\right)}\,.
\end{equation}

\subsection{Closed universes}

In the case of closed universes, the scale factor at the horizon crossing obtained from Eq.~\eqref{eq: friedmann 1} reads
\begin{equation}
    a_\star=\frac{\sqrt{n^2+K}}{\mathcal H_\Lambda}\,.
\end{equation}
Using the latter to invert Eq.~\eqref{eq: closedINF}, we find the conformal time at the horizon crossing:
\begin{equation}
 \tau_\star= -\dfrac{1}{\sqrt{K}}\arctan\left(\frac{\sqrt{K }}{n}\right).
\end{equation}
Thus, in view of Eqs.~\eqref{eq: rescaling inflation} and \eqref{eq:gen_sol_closed}, from Eq.~\eqref{eq: PPS2} we obtain the primordial power spectrum as
\begin{equation}\label{eq: ppsClosed}
    \mathcal{P}_T(n)=\left(\dfrac{\mathcal H_\Lambda}{\pi}\right)^2\frac{n^4}{n^4-K ^2}\,.
\end{equation}
This result is formally equivalent to Eq.~\eqref{eq:PS open}. Indeed, as for negative spatial curvatures, the effect of positive curvatures produces an enhancement of the primordial power spectrum compared to the flat universe. As $n\gg K$, the curvature effects become vanishingly small and one recovers the scale-invariant feature of the flat spectrum (cf. Eq.~\eqref{eq: PPS flat}).

We move now to study the evolution of the transfer functions in different epochs. As already discussed, we may start from the results obtained in the case of open universes and apply to them the transformations $\tau \rightarrow i \tau$ and $n\rightarrow i n$. For the radiation era, from Eq.~\eqref{eq:TF_RD_open} we find
\begin{equation}
\mathscr{T}_\text{rad}=\frac{\sqrt{K } \sin (n \tau)}{n \sin \left(\tau \sqrt{K } \right)}\,.
\label{eq:TF_RD_closed}
\end{equation}

Moreover, applying the same transformations to Eq.~\eqref{eq:TF_MD_open}, provides us with the transfer function in the matter era:
\begin{equation}
    \mathscr{T}_\text{mat}=\frac{6 K  n \cos (n \tau)-3 K ^{3/2} \sin (n \tau) \cot \left(\frac{ \tau \sqrt{K } }{2}\right)}{n \left(4 n^2-K \right) \left[\cos \left( \tau \sqrt{K }\right)-1\right]}\,.
    \label{eq:TF_MD_closed}
\end{equation}

In order to match the transfer functions in the radiation  and matter epochs, we consider the evolution of the radiation modes in the matter era. For this purpose, we consider the general solution in the intermediate regime obtained from Eq.~\eqref{eq:gen_sol_IR_open}:
\begin{align}
    &\mathscr{T}_\text{int}= \frac{A_\text{eq} \left[K ^{3/2} \cos (n \tau) \cot \left(\frac{\tau\sqrt{K }}{2}\right)+2 K  n \sin (n \tau)\right]}{\cos \left(\tau\sqrt{K }\right)-1} \nonumber \\
    &+ \frac{B_\text{eq} \left[2 K  n \cos (n \tau)-K ^{3/2} \sin (n \tau) \cot \left( \frac{\tau\sqrt{K }}{2}\right)\right]}{\cos \left(\tau\sqrt{K }\right)-1}\,.
\end{align}
Here, the coefficients $A_\text{eq}$ and $B_\text{eq}$ are to be fixed by matching Eq.~\eqref{eq:TF_RD_closed} with Eq.~\eqref{eq:TF_MD_closed}, and their respective derivatives, at the time of equivalence. Specifically, we find
\begin{align}
    &A_\text{eq}=\frac{2 n^2 \cot \left(\tau_\text{eq}\sqrt{K }\right)+\frac{2 K  \sin ^2(n \tau_\text{eq})-2 n^2}{\sin \left(\tau_\text{eq}\sqrt{K } \right)}-\frac{\sqrt{K } n \sin (2 n \tau_\text{eq})}{\cos \left(\tau_\text{eq}\sqrt{K } \right)+1}}{\sqrt{K } n^2 \left(4 n^2-K \right)} \\
   &B_\text{eq}=\dfrac{n+\dfrac{\sqrt{K } \sin (2 n \tau_\text{eq})}{\sin(\tau_\text{eq} \sqrt{K}) } +\dfrac{n \sin ^2(n \tau_\text{eq})}{\cos^2(\tau_\text{eq} \sqrt{K }/2)}}{n^2(4n^2-K)}\,.
\end{align}

\section{The Impact of effective Degrees of Freedom}
\label{sec:effective}

Although, in the standard cosmological picture, the evolution of the energy density in the radiation era is described by $\rho_\text{rad}\propto a^{-4}$,
such a behavior is not overall valid if one takes into account the intrinsic nature of particles that become nonrelativistic in different moments. Assuming the Universe to be an adiabatic system, as the temperature drops in view of the cosmic expansion, certain particles stop contributing to the radiation density before others. 

To analyze the effects of this phenomenon on the primordial power spectrum, we consider the Universe as a plasma composed of relativistic particles and photons in thermal equilibrium. Then, the radiation energy density and pressure can be expressed in terms of the plasma temperature $T$ as, respectively,
\begin{equation}
    \rho_\text{rad}(T)=\dfrac{\pi^2}{30}g_{\text{eff},\rho}(T)T^4\,, \quad p_\text{rad}(T)=\dfrac{\rho_\text{rad}(T)}{3}\,,
    \label{eq:radiation density}
\end{equation}
where $g_{\text{eff},\rho}$ is the effective number of relativistic degrees of freedom for species with mass $m_i\ll T_i$, contributing to the radiation energy density:
\begin{equation}
    g_{\text{eff},\rho}(T)=\sum_{i=\rm bosons}g_i\left(\dfrac{T_i}{T}\right)^4+\dfrac{7}{8}\sum_{i=\rm fermions}g_i\left(\dfrac{T_i}{T}\right)^4.
\end{equation}

\begin{figure}
    \begin{center}
    \includegraphics[width=3.4in]{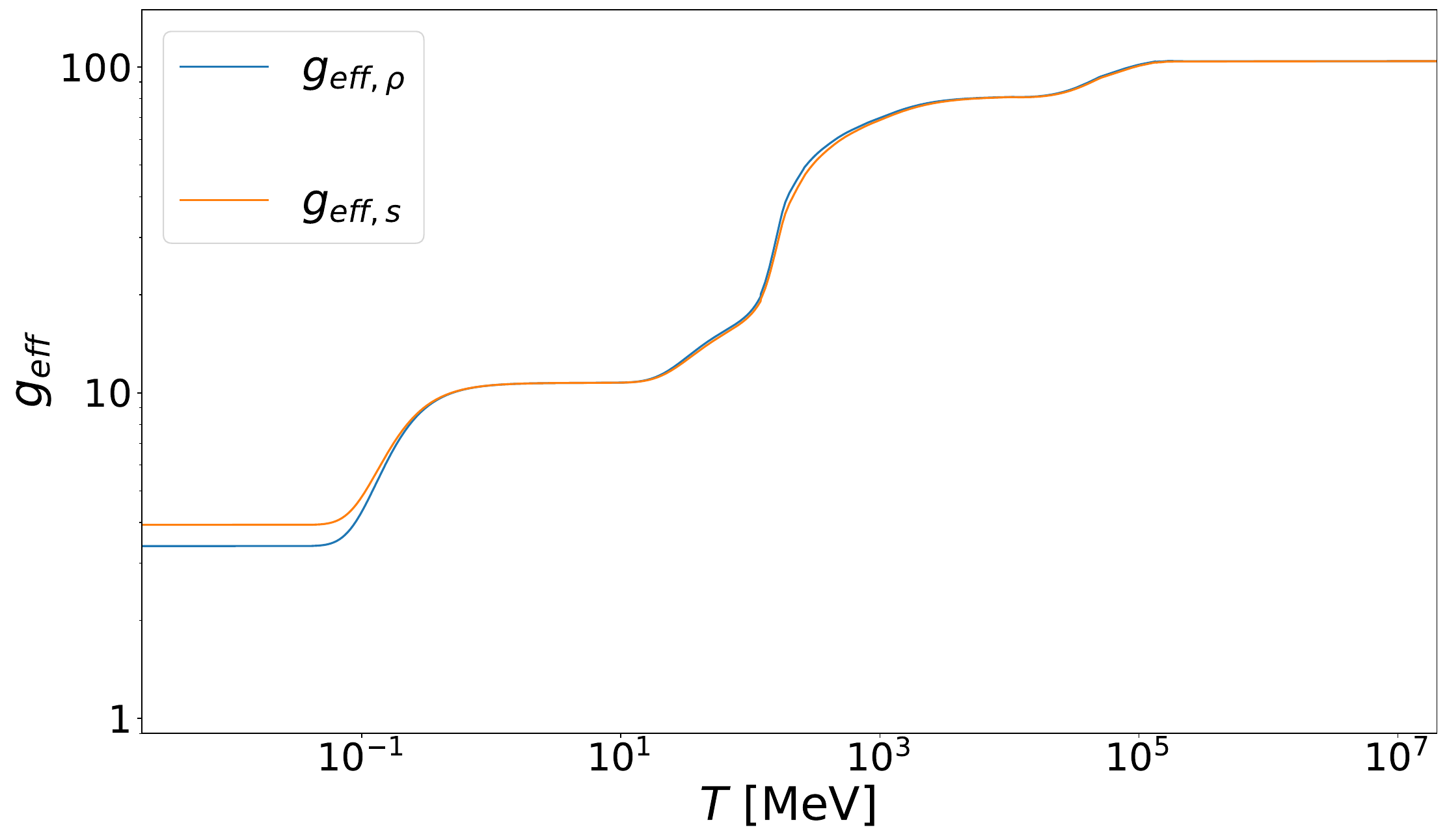}
    \caption{Evolution of the effective relativistic degrees of freedom as a function of the temperature of the Universe. Logarithmic scales are used on both axes.}
     \label{fig:geff}
    \end{center}
\end{figure}

Moreover, the laws of thermodynamics applied to an adiabatic system imply  the conservation of the entropy per comoving volume, $S=(\rho+p)a^3/T$. Thus, one has
\begin{equation}
    S(T)=s(T)a^3=\text{const}\,,
    \label{eq:entropy conservation}
\end{equation}
where $s$ is the entropy energy density that, to a very good approximation, is given by \cite{Kolb:1990vq}
\begin{equation}
    s(T)=\dfrac{2\pi^2}{45}g_{\text{eff},s} T^3\,.
    \label{eq:entropy density}
\end{equation}
Here, $g_{\text{eff},s}$ is the effective number of relativistic degrees of freedom contributing to the entropy:
\begin{equation}
    g_{\text{eff},s}(T)=\sum_{i=\rm bosons}g_i\left(\dfrac{T_i}{T}\right)^3+\dfrac{7}{8}\sum_{i=\rm fermions}g_i\left(\dfrac{T_i}{T}\right)^3.
\end{equation}
In Fig.~\ref{fig:geff}, we show the behaviors of $g_{\text{eff},\rho}(T)$ and $g_{\text{eff},s}(T)$, based on the data provided in Ref.~\cite{Saikawa:2018rcs}. The temperature evolution of the effective degrees of freedom depends on some peculiar transitions in cosmic history. For instance, $T\sim 1$ MeV marks the electron-positron annihilation, which occurs when the temperature of the Universe becomes of the order of the electron mass. Then, the most drastic change takes place at the quantum chromodynamics (QCD) phase transition, including the hadronic phase ($T< 100$ MeV), the nonperturbative phase ($T\sim 100$ MeV), and the perturbative phase ($T\gg 100$ MeV). Finally, another significant change is due to the electroweak crossover, when the temperature of the Universe transited from a high value of the symmetric phase to a low value of the broken phase regime characterized by a nonvanishing expectation value of the Higgs field \cite{Laine:2006cp}.
It is worth noting that $g_{\text{eff},s}$ can be identified with $g_{\text{eff},\rho}$ for most of the Universe's history, when all particle species shared the same temperature. However, in general, the two quantities cannot always be used interchangeably. 

Therefore, combining Eqs.~\eqref{eq:radiation density} and \eqref{eq:entropy conservation}  provides the correct evolution of the energy density during the radiation era:
\begin{equation}
    \rho_\text{rad} \propto g_{\text{eff},\rho}\, g_{\text{eff},s}^{-4/3}\,  a^{-4} \,.
    \label{eq: energy density modified}
\end{equation}
The latter suggests that deviations from the standard behavior occur unless $g_{\text{eff},\rho}$ and  $g_{\text{eff},s}$ are time independent.

In order to estimate the amplitude of GWs at the present time, we focus on the modes that entered the horizon while the Universe was still in the radiation era. An appropriate description of the evolution of the transfer function after the modes reenter the horizon may be given by the WKB approximation, which is valid for wavelengths that are much shorter than the cosmic transition through the matter era. Thus,  $\mathscr{T}\propto a^{-1}e^{\pm i n \tau}$, implying \cite{Saikawa:2018rcs}
\begin{equation}
    \mathscr{T}^{\prime}(\tau,n)^2\approx \dfrac{n^2}{2}\left[\dfrac{a_\star}{a(\tau)}\right]^2=\dfrac{\mathcal{H}_\star^2\, a_\star^2}{2a(\tau)^2}\,,
\end{equation}
where $\mathcal{H}_\star\equiv\mathcal{H}(\tau_\star)=n$. Hence, from Eq.~\eqref{eq: spectral energy density}, we find
\begin{equation}
    \Omega_\text{GW}(\tau,n)=\dfrac{\mathcal{P}_T(n)}{24}\left[\dfrac{\mathcal{H}_\star \, a_\star}{\mathcal{H}(\tau)a(\tau)}\right]^2\,.
    \label{eq: spectral energy density species}
\end{equation}
If the horizon crossing occurs in the radiation era, then we have
\begin{equation}
    \left(\frac{\mathcal{H}_\star}{\mathcal{H}_0}\right)^2= \frac{\pi^2}{30}\frac{g_{\text{eff},\rho\star}\,T_\star^4\, a_\star^2}{\rho_\text{cr,0}}-\dfrac{K}{\mathcal{H}_0^2}\,,
    \label{eq: H hc}
\end{equation}
where $g_{\text{eff},\rho\star}\equiv g_{\text{eff},\rho}(T_\star)$, with $T_\star$ being the temperature of the Universe at the horizon crossing.
The above equation can be manipulated by introducing the curvature density fraction $\Omega_{K,0}\equiv -K/\mathcal{H}_0^2$ and the
photon density fraction $\Omega_{\gamma,0}\equiv \rho_{\gamma,0}/\rho_\text{cr,0}$, where $\rho_{\gamma,0}=\frac{\pi^2}{15}T_{\gamma,0}^4$ is the present-day value of the photon energy density and $T_{\gamma,0}$ the corresponding temperature. In addition, from Eqs.~\eqref{eq:entropy conservation} and \eqref{eq:entropy density}, it follows that
\begin{equation}
    g_{\text{eff},s\star}\, a_\star^3\,  T_\star^3=g_{\text{eff},s0}\,T_{\gamma,0}^3\,,
    \label{eq:conservation}
\end{equation}
where $g_{\text{eff},s0}\equiv g_{\text{eff},s}(T_{\gamma,0})$.
Therefore, Eq.~\eqref{eq: H hc} becomes
\begin{equation}
    \left(\frac{\mathcal{H}_\star}{\mathcal{H}_0}\right)^2=\frac{\Omega_{\gamma,0}}{2a_\star^2}g_{\text{eff},\rho\star}\left(\frac{g_{\text{eff},s0}}{g_{\text{eff},s\star}}\right)^{4/3}+\Omega_{K,0}\,.
    \label{eq: H hc 2}
\end{equation}
In view of Eqs.~\eqref{eq:conservation} and \eqref{eq: H hc 2}, Eq.~\eqref{eq: spectral energy density species} evaluated at the present time  yields
\begin{align} \label{eq: spectrum + suppres}
  &\Omega_\text{GW}(\tau_0,n)= \frac{\mathcal P_T(n)}{48}\left(\frac{g_{\text{eff},s0}}{g_{\text{eff},s\star}}\right)^{2/3} \times  \nonumber \\
  &\times \left[\Omega_{\gamma,0}\, g_{\text{eff},\rho\star}\left(\frac{g_{\text{eff},s0}}{g_{\text{eff},s\star}}\right)^{2/3} + 2\Omega_{K,0}\left(  \dfrac{T_{\gamma,0}}{T_\star}\right)^2\right]. 
\end{align}
The last term in the square bracket shows the influence of nonvanishing spatial curvature on the current value of the spectral energy density of GWs. We note that, for $\Omega_{K,0}=0$, we recover the expression for a flat universe provided in Ref.~\cite{Saikawa:2018rcs}.

Finally, one could relate the frequency of GWs to the horizon crossing of the corresponding mode:
\begin{align}
    f&=\dfrac{n}{2\pi}= \dfrac{\mathcal{H}_\star}{2    \pi}=   \nonumber  \\
    &=\dfrac{\mathcal{H}_0}{2\pi}\sqrt{\frac{\Omega_{\gamma,0}}{2}\left(\frac{g_{\text{eff},s0}}{g_{\text{eff},s\star}}\right)^{2/3} g_{\text{eff},\rho\star} \left(\dfrac{T_\star}{T_{\gamma,0}}\right)^2+ \Omega_{K,0}}\,.
\end{align}
\begin{figure}
    \begin{center}
    \includegraphics[width=3.4in]{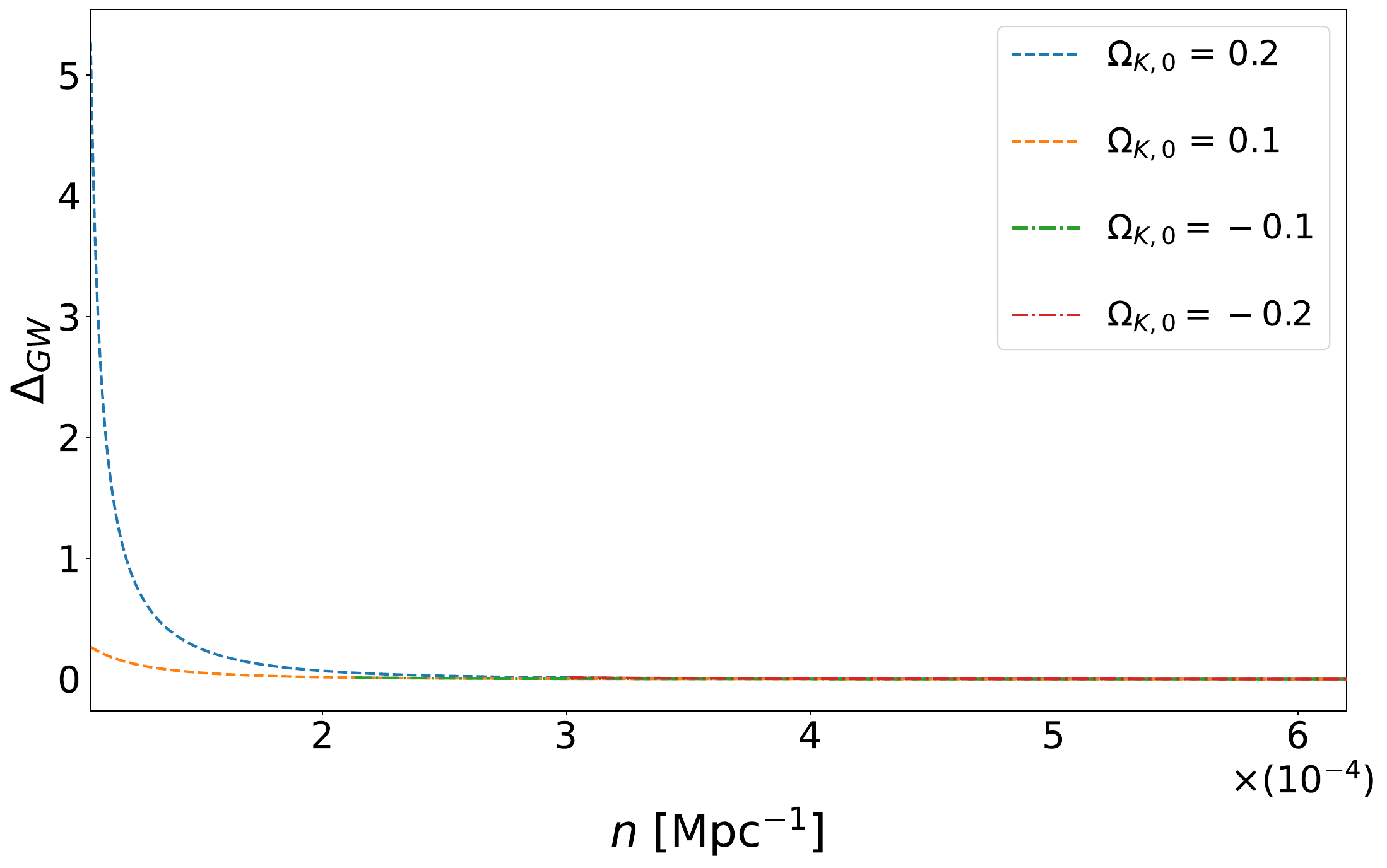}
    \caption{Relative change of the primordial power spectrum of GWs for different spatial curvatures with respect to the flat universe case.}
     \label{fig:residuals}
    \end{center}
\end{figure}
\begin{figure*}
    \centering
    \includegraphics[width=\textwidth]{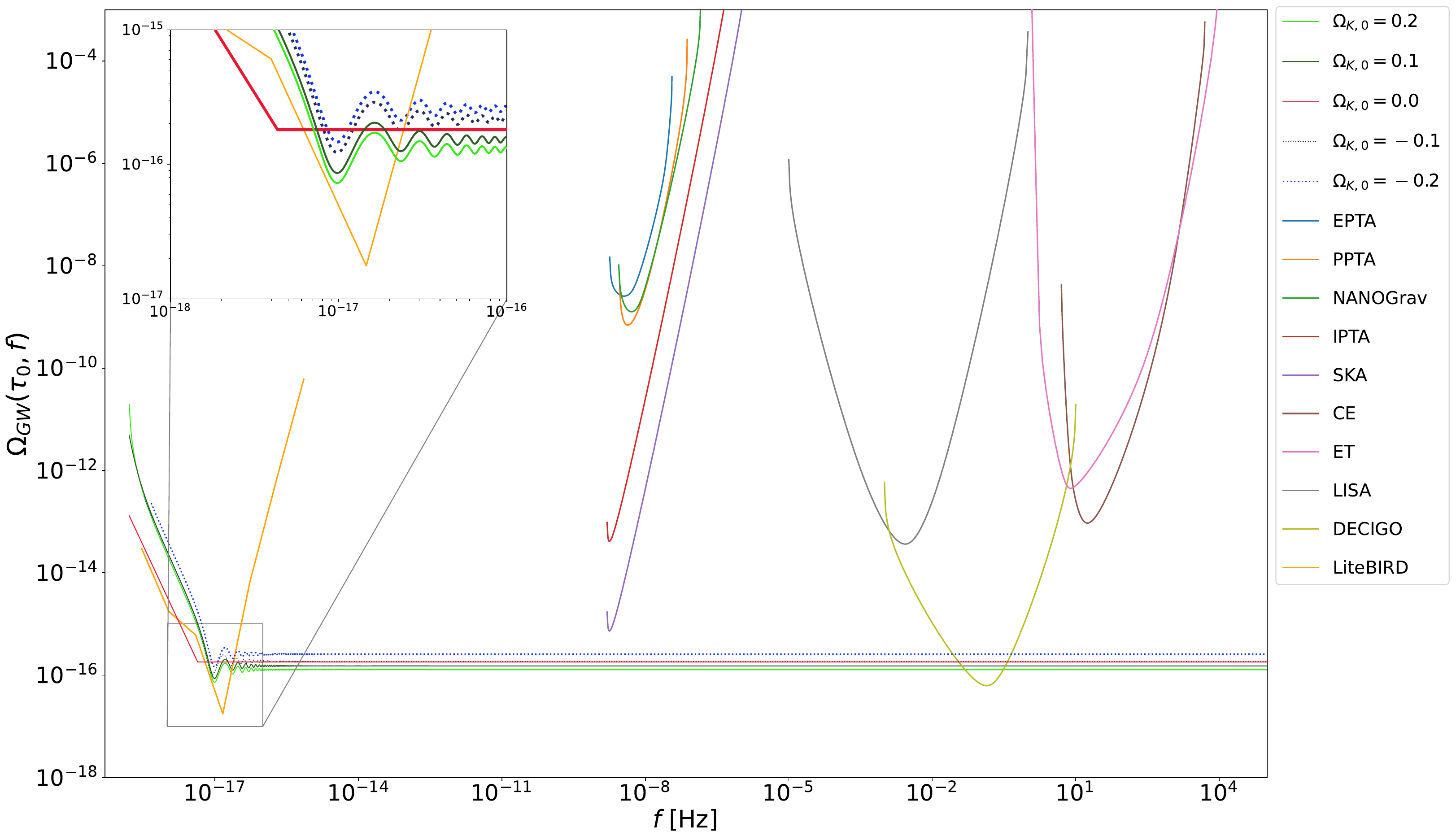}
    \caption{Average spectral energy density of GWs, at the present time, for different spatial curvatures, together with the PLI sensitivity curves of present and future GW experiments (see Ref.~\cite{Schmitz:2020syl} and references therein). The corrections of the effective degrees of freedom of relativistic species are here neglected. Logarithmic scales are used on both axes.}
     \label{fig:spectrum}
\end{figure*}

\section{Observational consequences}
\label{sec:observations}

We here investigate the observational consequences of the results obtained in the previous section. In particular, to estimate the impact of the spatial curvature on the primordial power spectrum, we consider the quantity
\begin{equation}
    \Delta_\text{GW}\equiv \dfrac{|\mathcal{P}_T^\text{(curved)}-\mathcal{P}_T^\text{(flat)}|}{\mathcal{P}_T^\text{(flat)}}\,,
\end{equation}
accounting for the relative differences with respect to the flat case. 
We show the results in terms of different $\Omega_{K,0}$ values in Fig.~\ref{fig:residuals}, where we fixed the Hubble constant to the latest estimate by the Planck collaboration, $\mathcal{H}_0=67.4$ km/s/Mpc \cite{Planck:2018vyg}.
We note that significant differences occur at low wavenumbers. Although the symmetric form of the spectrum for open and closed universes, it is important to bear in mind the different domains of the generalized wavenumber in the two cases (see Sec.~\ref{sec:perturbations}).
Moreover, one could relate the stochastic background of GWs to the sensitivity response of detectors \cite{Thrane:2013,Alonso:2020rar,Schmitz:2020syl,DEramo:2019tit}.
The observed signal, $s_I(f)$, may be written as the sum of the strain induced by the GW signal, $h_I(f)$, and the $I$th detector noise, $n_I(f)$:
\begin{equation}
    s_I(f) = h_I(f) + n_I(f)\, .
\end{equation}
Assuming both GW background and noise to be Gaussian and stationary, we define the strain power spectral density of a GW background ($S_{h}$) and the detector power spectral density due to the noise ($P_{I}$) as, respectively,
\begin{align}
    \langle h_I(f)h_J^{*}(f')\rangle = \frac{1}{2}\delta(f-f') \Gamma_{IJ}S_{h}\, , \\
    \langle n_I(f)n_J^{*}(f')\rangle = \frac{1}{2}\delta(f-f') \delta_{IJ}P_{I}\,,
\end{align}
where $\Gamma_{IJ}$ is the overlap reduction function for the correlated response between the detectors $I$ and $J$ to a GW background \cite{Thrane:2013}. Such a function takes into account the different locations and orientations of the detectors.

Furthermore, the signal-to-noise ratio (SNR) for a cross-correlation search for an unpolarized and isotropic stochastic background is given by
\begin{equation}\label{eq: SNR background}
    \text{SNR} = \sqrt{2\mathcal{T}}\left[\int_{f_{min}}^{f_{max}}df\ \sum_{I=1}^{M}\sum_{J>I}^{M} \frac{\Gamma_{IJ}^{2}S_{h}^2(f)}{P_{I}(f)P_{J}(f)}\right]^{\frac{1}{2}} ,
\end{equation}
where $M$ labels the number of detectors and $\mathcal{T}$ is the observational time.
Starting from Eq.~\eqref{eq: SNR background}, we write the effective strain noise as
\begin{equation}\label{eq: Seff}
    S_\text{eff}(f)=\left[ \sum_{I=1}^{M}\sum_{J>I}^{M} \frac{\Gamma_{IJ}^{2}S_{h}^2(f)}{P_{I}(f)P_{J}(f)}\right]^{-\frac{1}{2}}\,,
\end{equation}
which could be expressed as a function of the energy density units $\Omega_\text{eff}(f)$:
\begin{equation}
  \Omega_\text{eff}(f) =\frac{2 \pi^2}{3 \mathcal{H}_0^2}f^3 S_\text{eff}(f)\,.
\end{equation}
Then, one can calculate the amplitude $\Omega_{\xi}$ such for the integrated SNR to have a fixed value. This is given by
\begin{equation}
    \Omega_{\xi}=\frac{\text{SNR}}{\sqrt{2\mathcal{T}}}\left[\int_{f_\text{min}}^{f_\text{max}}df\ \frac{(f/ f_\text{ref})^{2\xi}}{\Omega_\text{eff}^{2}(f)} \right]^{-\frac{1}{2}}\, ,
\end{equation}
where $f_\text{ref}$ is a reference frequency, and $\xi$ is the so-called spectral index. It is worth stressing that the choice of $f_\text{ref}$ is arbitrary and will not affect the final result. 
Finally, the power-law integrated (PLI) sensitivity curve is defined as
\begin{equation}
    \Omega_\text{PLI}(f)=\max_{\xi} \left[ \Omega_{\xi}\left( \frac{f}{f_\text{ref}}\right)^{\xi} \right]\, .
\end{equation}

In Fig.~\ref{fig:spectrum}, we display the average over oscillations of the spectral energy density of GWs for different $\Omega_{K,0}$ values, where we neglect the corrections due to the effective number of relativistic degrees of freedom. 
For reference, we also show the PLI detector curves for $\text{SNR}=1$ and the following observation times: $\mathcal{T}=1$ yr for ET, CE, LISA and DECIGO; $\mathcal{T}=11$ yrs for PPTA; $\mathcal{T}=12$ yrs for NANOGrav; $\mathcal{T}=18$ yrs for EPTA; $\mathcal{T}=20$ yrs for SKA and IPTA \cite{Schmitz:2020syl}.
Moreover, we include the sensitivity curve expected for the LiteBIRD experiment, aiming at the detection of the CMB $B$ modes \cite{Campeti:2020xwn}.
We notice that the most promising detector for observing the GW signal at high frequencies is the DECIGO experiment. On the other hand, the LiteBIRD experiment will be the most sensitive probe to detect the low-frequency signal of GWs through the measurement of the CMB polarization patterns.

\begin{figure*}
    \centering
    \includegraphics[width=0.9\textwidth]{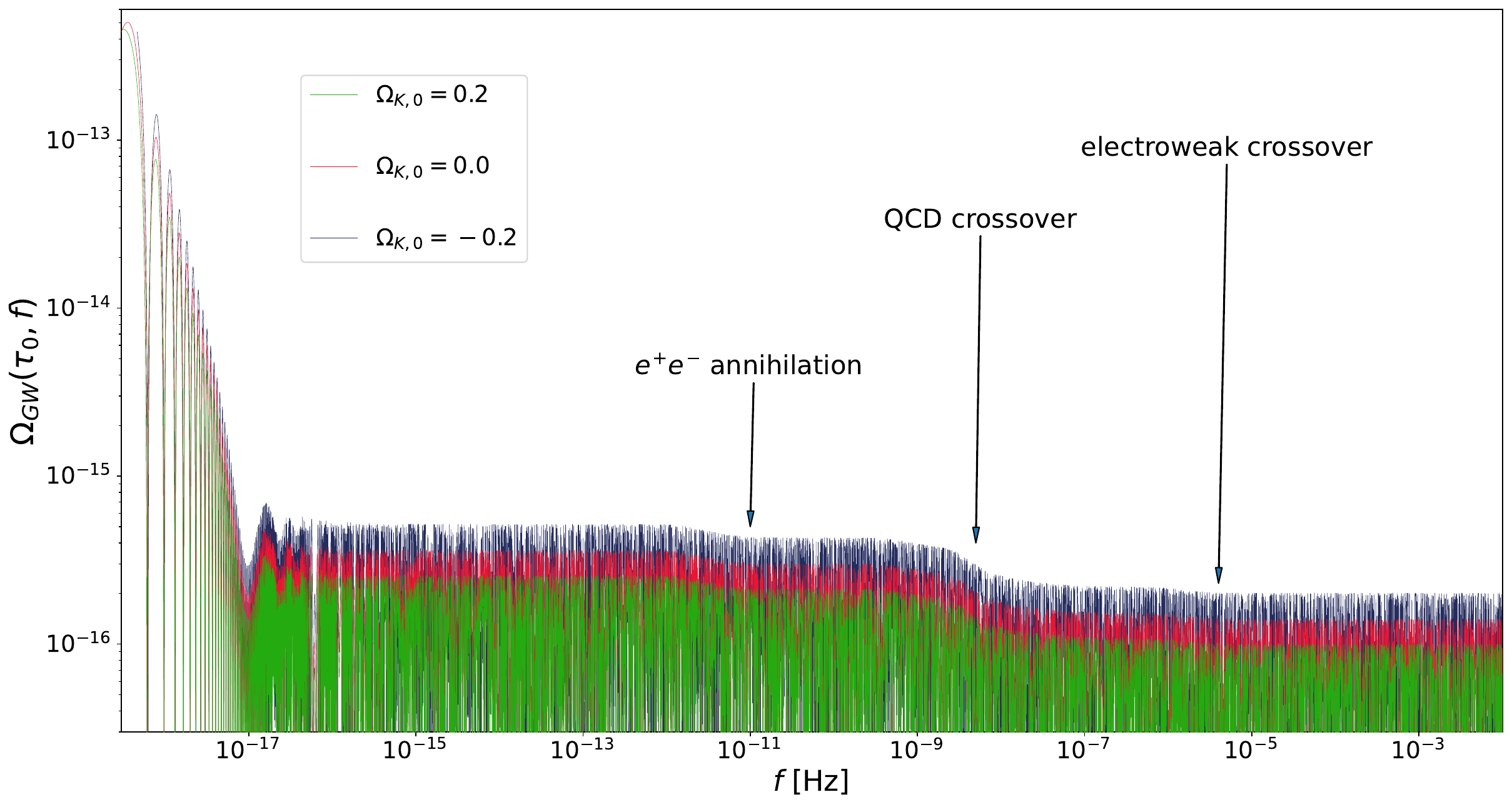}
    \caption{Present-time spectral energy density of GWs for different spatial curvatures, taking into account the evolution of the effective number of relativistic degrees of freedom. The different phase transitions across cosmic history are specified in correspondence to the largest suppressions of the spectrum. Logarithmic scales are used on both axes.}
    \label{fig:spectrum+suppres}
\end{figure*}

Moreover, in Fig.~\ref{fig:spectrum+suppres}, we highlight the effects due to the evolving degrees of freedom of relativistic species on the spectral energy density of GWs for 20\% deviations from the flat case. We can see that the amplitude of GWs is suppressed at the frequencies 
of particular phase transitions occurring in the early Universe, such that the electron-positron annihilation, the QCD phase transition and the electroweak crossover. 

\section{Conclusions}
\label{sec:conclusions}

In this work, we analyzed the role of nonvanishing spatial curvature on the primordial power spectrum of GWs. In the framework of standard FLRW cosmology, we solved the Friedmann equations in the de Sitter inflationary stage, as well as in the radiation and matter epochs, in the presence of a nonzero curvature parameter.  Afterward, we introduced linear perturbations over the spatial part of the background metric, in order to study the propagation of GWs during inflation. We thus characterized the spectrum of tensor harmonics and derived the normalization conditions for the GW amplitude by means of creation and annihilation quantum operators. 
Hence, we analytically solved the dynamics of tensor perturbations for flat, open and closed universes under the Bunch-Davies vacuum, which allowed us to fix the arbitrary constants of integration resulting from the general solutions of the GW propagation equation.

Then, we framed the evolution of tensor modes in terms of the amplitude of GWs that left the horizon during inflation, and the transfer functions accounting for the GW modes evolution in the post-inflationary era. By distinguishing the cases of zero, negative and positive spatial curvatures, we used the background cosmological solutions to obtain the power spectrum of primordial GWs at the time of the horizon crossing. 
Also, we found the expressions of the transfer functions in the radiation and matter eras, and analyzed the propagation of radiation modes into the matter era to smoothly match the solutions between the two epochs.

Furthermore, we investigated the corrections to the standard cosmic behavior of the radiation energy density due to the effective  degrees of freedom of relativistic species.
In particular, assuming the Universe to be an adiabatic system and the cosmic fluid as a hot plasma made of relativistic particles in thermal equilibrium with photons, we studied the temperature evolution of the effective number of relativistic degrees of freedom contributing to the radiation and entropy energy densities. 
Consequently, we showed the impact of these on the relic energy density of GWs. Specifically, we found that the primordial power spectrum presents major suppressions at frequencies that correspond to three phase transitions through cosmic history, namely the electron-positron annihilation, the QCD crossover and the electroweak crossover. These features appear to be common to all FLRW models, regardless of the value of spatial curvature, whose variation produces an overall shift in the amplitude of the GW spectrum.

Finally, we discussed the observational consequences of our theoretical predictions. In particular, we described how to relate the stochastic background of GWs to the sensitivity response of a given detector. We thus introduced the SNR in terms of the strain power spectral density of a GW background and the detector noise. Then, we computed the average spectral energy density of GWs for different values of the curvature parameter, and compared them with the PLI sensitivity curves of present and upcoming GW experiments. 
In particular, we found that the future measurements of the LiteBIRD satellite represent the most sensitive probe to distinguish among different spatial curvatures in the low-frequency regime of the primordial GW signal.

The spatial curvature effects on tensor perturbations may be further explored by including a scalar field in the gravitational action \cite{Chowdhury:2022gdc} or, also, by considering different formulations of the gravitational interaction in the context of modified theories of gravity \cite{DAgostino:2022tdk}. Future investigations will be dedicated to this scope.

\acknowledgements

The Authors are grateful to the Istituto Nazionale di Fisica Nucleare (INFN) -- Sezione di Napoli, {\it iniziative specifiche} QGSKY, MOONLIGHT and TEONGRAV. 
D.V. acknowledges the FCT Project No. PTDC/FIS-AST/0054/2021.

\bibliography{biblio}

\begin{thebibliography}{80}%
\makeatletter
\providecommand \@ifxundefined [1]{%
 \@ifx{#1\undefined}
}%
\providecommand \@ifnum [1]{%
 \ifnum #1\expandafter \@firstoftwo
 \else \expandafter \@secondoftwo
 \fi
}%
\providecommand \@ifx [1]{%
 \ifx #1\expandafter \@firstoftwo
 \else \expandafter \@secondoftwo
 \fi
}%
\providecommand \natexlab [1]{#1}%
\providecommand \enquote  [1]{``#1''}%
\providecommand \bibnamefont  [1]{#1}%
\providecommand \bibfnamefont [1]{#1}%
\providecommand \citenamefont [1]{#1}%
\providecommand \href@noop [0]{\@secondoftwo}%
\providecommand \href [0]{\begingroup \@sanitize@url \@href}%
\providecommand \@href[1]{\@@startlink{#1}\@@href}%
\providecommand \@@href[1]{\endgroup#1\@@endlink}%
\providecommand \@sanitize@url [0]{\catcode `\\12\catcode `\$12\catcode
  `\&12\catcode `\#12\catcode `\^12\catcode `\_12\catcode `\%12\relax}%
\providecommand \@@startlink[1]{}%
\providecommand \@@endlink[0]{}%
\providecommand \url  [0]{\begingroup\@sanitize@url \@url }%
\providecommand \@url [1]{\endgroup\@href {#1}{\urlprefix }}%
\providecommand \urlprefix  [0]{URL }%
\providecommand \Eprint [0]{\href }%
\providecommand \doibase [0]{http://dx.doi.org/}%
\providecommand \selectlanguage [0]{\@gobble}%
\providecommand \bibinfo  [0]{\@secondoftwo}%
\providecommand \bibfield  [0]{\@secondoftwo}%
\providecommand \translation [1]{[#1]}%
\providecommand \BibitemOpen [0]{}%
\providecommand \bibitemStop [0]{}%
\providecommand \bibitemNoStop [0]{.\EOS\space}%
\providecommand \EOS [0]{\spacefactor3000\relax}%
\providecommand \BibitemShut  [1]{\csname bibitem#1\endcsname}%
\let\auto@bib@innerbib\@empty
\bibitem [{\citenamefont {Aghanim}\ \emph {et~al.}(2020)\citenamefont {Aghanim}
  \emph {et~al.}}]{Planck:2018vyg}%
  \BibitemOpen
  \bibfield  {author} {\bibinfo {author} {\bibfnamefont {N.}~\bibnamefont
  {Aghanim}} \emph {et~al.} (\bibinfo {collaboration} {Planck Collaboration}),\
  }\href {\doibase 10.1051/0004-6361/201833910} {\bibfield  {journal} {\bibinfo
   {journal} {Astron. Astrophys.}\ }\textbf {\bibinfo {volume} {641}},\
  \bibinfo {pages} {A6} (\bibinfo {year} {2020})},\ \Eprint
  {http://arxiv.org/abs/1807.06209} {arXiv:1807.06209 [astro-ph.CO]}
  \BibitemShut {NoStop}%
\bibitem [{\citenamefont {Riess}\ \emph {et~al.}(1998)\citenamefont {Riess}
  \emph {et~al.}}]{SupernovaSearchTeam:1998fmf}%
  \BibitemOpen
  \bibfield  {author} {\bibinfo {author} {\bibfnamefont {A.~G.}\ \bibnamefont
  {Riess}} \emph {et~al.} (\bibinfo {collaboration} {Supernova Search Team}),\
  }\href {\doibase 10.1086/300499} {\bibfield  {journal} {\bibinfo  {journal}
  {Astron. J.}\ }\textbf {\bibinfo {volume} {116}},\ \bibinfo {pages} {1009}
  (\bibinfo {year} {1998})},\ \Eprint {http://arxiv.org/abs/astro-ph/9805201}
  {arXiv:astro-ph/9805201} \BibitemShut {NoStop}%
\bibitem [{\citenamefont {Perlmutter}\ \emph {et~al.}(1999)\citenamefont
  {Perlmutter} \emph {et~al.}}]{SupernovaCosmologyProject:1998vns}%
  \BibitemOpen
  \bibfield  {author} {\bibinfo {author} {\bibfnamefont {S.}~\bibnamefont
  {Perlmutter}} \emph {et~al.} (\bibinfo {collaboration} {Supernova Cosmology
  Project}),\ }\href {\doibase 10.1086/307221} {\bibfield  {journal} {\bibinfo
  {journal} {Astrophys. J.}\ }\textbf {\bibinfo {volume} {517}},\ \bibinfo
  {pages} {565} (\bibinfo {year} {1999})},\ \Eprint
  {http://arxiv.org/abs/astro-ph/9812133} {arXiv:astro-ph/9812133} \BibitemShut
  {NoStop}%
\bibitem [{\citenamefont {Carroll}(2001)}]{Carroll:2000fy}%
  \BibitemOpen
  \bibfield  {author} {\bibinfo {author} {\bibfnamefont {S.~M.}\ \bibnamefont
  {Carroll}},\ }\href {\doibase 10.12942/lrr-2001-1} {\bibfield  {journal}
  {\bibinfo  {journal} {Living Rev. Rel.}\ }\textbf {\bibinfo {volume} {4}},\
  \bibinfo {pages} {1} (\bibinfo {year} {2001})},\ \Eprint
  {http://arxiv.org/abs/astro-ph/0004075} {arXiv:astro-ph/0004075} \BibitemShut
  {NoStop}%
\bibitem [{\citenamefont {Peebles}\ and\ \citenamefont
  {Ratra}(2003)}]{Peebles:2002gy}%
  \BibitemOpen
  \bibfield  {author} {\bibinfo {author} {\bibfnamefont {P.~J.~E.}\
  \bibnamefont {Peebles}}\ and\ \bibinfo {author} {\bibfnamefont
  {B.}~\bibnamefont {Ratra}},\ }\href {\doibase 10.1103/RevModPhys.75.559}
  {\bibfield  {journal} {\bibinfo  {journal} {Rev. Mod. Phys.}\ }\textbf
  {\bibinfo {volume} {75}},\ \bibinfo {pages} {559} (\bibinfo {year} {2003})},\
  \Eprint {http://arxiv.org/abs/astro-ph/0207347} {arXiv:astro-ph/0207347}
  \BibitemShut {NoStop}%
\bibitem [{\citenamefont {D'Agostino}\ \emph {et~al.}(2022)\citenamefont
  {D'Agostino}, \citenamefont {Luongo},\ and\ \citenamefont
  {Muccino}}]{DAgostino:2022fcx}%
  \BibitemOpen
  \bibfield  {author} {\bibinfo {author} {\bibfnamefont {R.}~\bibnamefont
  {D'Agostino}}, \bibinfo {author} {\bibfnamefont {O.}~\bibnamefont {Luongo}},
  \ and\ \bibinfo {author} {\bibfnamefont {M.}~\bibnamefont {Muccino}},\ }\href
  {\doibase 10.1088/1361-6382/ac8af2} {\bibfield  {journal} {\bibinfo
  {journal} {Class. Quant. Grav.}\ }\textbf {\bibinfo {volume} {39}},\ \bibinfo
  {pages} {195014} (\bibinfo {year} {2022})},\ \Eprint
  {http://arxiv.org/abs/2204.02190} {arXiv:2204.02190 [gr-qc]} \BibitemShut
  {NoStop}%
\bibitem [{\citenamefont {Sotiriou}\ and\ \citenamefont
  {Faraoni}(2010)}]{Sotiriou:2008rp}%
  \BibitemOpen
  \bibfield  {author} {\bibinfo {author} {\bibfnamefont {T.~P.}\ \bibnamefont
  {Sotiriou}}\ and\ \bibinfo {author} {\bibfnamefont {V.}~\bibnamefont
  {Faraoni}},\ }\href {\doibase 10.1103/RevModPhys.82.451} {\bibfield
  {journal} {\bibinfo  {journal} {Rev. Mod. Phys.}\ }\textbf {\bibinfo {volume}
  {82}},\ \bibinfo {pages} {451} (\bibinfo {year} {2010})},\ \Eprint
  {http://arxiv.org/abs/0805.1726} {arXiv:0805.1726 [gr-qc]} \BibitemShut
  {NoStop}%
\bibitem [{\citenamefont {Linder}(2010)}]{Linder:2010py}%
  \BibitemOpen
  \bibfield  {author} {\bibinfo {author} {\bibfnamefont {E.~V.}\ \bibnamefont
  {Linder}},\ }\href {\doibase 10.1103/PhysRevD.81.127301} {\bibfield
  {journal} {\bibinfo  {journal} {Phys. Rev. D}\ }\textbf {\bibinfo {volume}
  {81}},\ \bibinfo {pages} {127301} (\bibinfo {year} {2010})},\ \Eprint
  {http://arxiv.org/abs/1005.3039} {arXiv:1005.3039 [astro-ph.CO]} \BibitemShut
  {NoStop}%
\bibitem [{\citenamefont {D'Agostino}\ and\ \citenamefont
  {Luongo}(2018)}]{DAgostino:2018ngy}%
  \BibitemOpen
  \bibfield  {author} {\bibinfo {author} {\bibfnamefont {R.}~\bibnamefont
  {D'Agostino}}\ and\ \bibinfo {author} {\bibfnamefont {O.}~\bibnamefont
  {Luongo}},\ }\href {\doibase 10.1103/PhysRevD.98.124013} {\bibfield
  {journal} {\bibinfo  {journal} {Phys. Rev. D}\ }\textbf {\bibinfo {volume}
  {98}},\ \bibinfo {pages} {124013} (\bibinfo {year} {2018})},\ \Eprint
  {http://arxiv.org/abs/1807.10167} {arXiv:1807.10167 [gr-qc]} \BibitemShut
  {NoStop}%
\bibitem [{\citenamefont {Jim\'enez}\ \emph {et~al.}(2020)\citenamefont
  {Jim\'enez}, \citenamefont {Heisenberg}, \citenamefont {Koivisto},\ and\
  \citenamefont {Pekar}}]{BeltranJimenez:2019tme}%
  \BibitemOpen
  \bibfield  {author} {\bibinfo {author} {\bibfnamefont {J.~B.}\ \bibnamefont
  {Jim\'enez}}, \bibinfo {author} {\bibfnamefont {L.}~\bibnamefont
  {Heisenberg}}, \bibinfo {author} {\bibfnamefont {T.}~\bibnamefont
  {Koivisto}}, \ and\ \bibinfo {author} {\bibfnamefont {S.}~\bibnamefont
  {Pekar}},\ }\href {\doibase 10.1103/PhysRevD.101.103507} {\bibfield
  {journal} {\bibinfo  {journal} {Phys. Rev. D}\ }\textbf {\bibinfo {volume}
  {101}},\ \bibinfo {pages} {103507} (\bibinfo {year} {2020})},\ \Eprint
  {http://arxiv.org/abs/1906.10027} {arXiv:1906.10027 [gr-qc]} \BibitemShut
  {NoStop}%
\bibitem [{\citenamefont {Capozziello}\ \emph {et~al.}(2022)\citenamefont
  {Capozziello}, \citenamefont {D'Agostino},\ and\ \citenamefont
  {Luongo}}]{Capozziello:2022rac}%
  \BibitemOpen
  \bibfield  {author} {\bibinfo {author} {\bibfnamefont {S.}~\bibnamefont
  {Capozziello}}, \bibinfo {author} {\bibfnamefont {R.}~\bibnamefont
  {D'Agostino}}, \ and\ \bibinfo {author} {\bibfnamefont {O.}~\bibnamefont
  {Luongo}},\ }\href {\doibase 10.1016/j.physletb.2022.137475} {\bibfield
  {journal} {\bibinfo  {journal} {Phys. Lett. B}\ }\textbf {\bibinfo {volume}
  {834}},\ \bibinfo {pages} {137475} (\bibinfo {year} {2022})},\ \Eprint
  {http://arxiv.org/abs/2207.01276} {arXiv:2207.01276 [gr-qc]} \BibitemShut
  {NoStop}%
\bibitem [{\citenamefont {Kamenshchik}\ \emph {et~al.}(2001)\citenamefont
  {Kamenshchik}, \citenamefont {Moschella},\ and\ \citenamefont
  {Pasquier}}]{Kamenshchik:2001cp}%
  \BibitemOpen
  \bibfield  {author} {\bibinfo {author} {\bibfnamefont {A.~Y.}\ \bibnamefont
  {Kamenshchik}}, \bibinfo {author} {\bibfnamefont {U.}~\bibnamefont
  {Moschella}}, \ and\ \bibinfo {author} {\bibfnamefont {V.}~\bibnamefont
  {Pasquier}},\ }\href {\doibase 10.1016/S0370-2693(01)00571-8} {\bibfield
  {journal} {\bibinfo  {journal} {Phys. Lett. B}\ }\textbf {\bibinfo {volume}
  {511}},\ \bibinfo {pages} {265} (\bibinfo {year} {2001})},\ \Eprint
  {http://arxiv.org/abs/gr-qc/0103004} {arXiv:gr-qc/0103004} \BibitemShut
  {NoStop}%
\bibitem [{\citenamefont {Scherrer}(2004)}]{Scherrer:2004au}%
  \BibitemOpen
  \bibfield  {author} {\bibinfo {author} {\bibfnamefont {R.~J.}\ \bibnamefont
  {Scherrer}},\ }\href {\doibase 10.1103/PhysRevLett.93.011301} {\bibfield
  {journal} {\bibinfo  {journal} {Phys. Rev. Lett.}\ }\textbf {\bibinfo
  {volume} {93}},\ \bibinfo {pages} {011301} (\bibinfo {year} {2004})},\
  \Eprint {http://arxiv.org/abs/astro-ph/0402316} {arXiv:astro-ph/0402316}
  \BibitemShut {NoStop}%
\bibitem [{\citenamefont {Capozziello}\ \emph {et~al.}(2018)\citenamefont
  {Capozziello}, \citenamefont {D'Agostino},\ and\ \citenamefont
  {Luongo}}]{Capozziello:2017buj}%
  \BibitemOpen
  \bibfield  {author} {\bibinfo {author} {\bibfnamefont {S.}~\bibnamefont
  {Capozziello}}, \bibinfo {author} {\bibfnamefont {R.}~\bibnamefont
  {D'Agostino}}, \ and\ \bibinfo {author} {\bibfnamefont {O.}~\bibnamefont
  {Luongo}},\ }\href {\doibase 10.1016/j.dark.2018.02.002} {\bibfield
  {journal} {\bibinfo  {journal} {Phys. Dark Univ.}\ }\textbf {\bibinfo
  {volume} {20}},\ \bibinfo {pages} {1} (\bibinfo {year} {2018})},\ \Eprint
  {http://arxiv.org/abs/1712.04317} {arXiv:1712.04317 [gr-qc]} \BibitemShut
  {NoStop}%
\bibitem [{\citenamefont {Li}(2004)}]{Li:2004rb}%
  \BibitemOpen
  \bibfield  {author} {\bibinfo {author} {\bibfnamefont {M.}~\bibnamefont
  {Li}},\ }\href {\doibase 10.1016/j.physletb.2004.10.014} {\bibfield
  {journal} {\bibinfo  {journal} {Phys. Lett. B}\ }\textbf {\bibinfo {volume}
  {603}},\ \bibinfo {pages} {1} (\bibinfo {year} {2004})},\ \Eprint
  {http://arxiv.org/abs/hep-th/0403127} {arXiv:hep-th/0403127} \BibitemShut
  {NoStop}%
\bibitem [{\citenamefont {D'Agostino}(2019)}]{DAgostino:2019wko}%
  \BibitemOpen
  \bibfield  {author} {\bibinfo {author} {\bibfnamefont {R.}~\bibnamefont
  {D'Agostino}},\ }\href {\doibase 10.1103/PhysRevD.99.103524} {\bibfield
  {journal} {\bibinfo  {journal} {Phys. Rev. D}\ }\textbf {\bibinfo {volume}
  {99}},\ \bibinfo {pages} {103524} (\bibinfo {year} {2019})},\ \Eprint
  {http://arxiv.org/abs/1903.03836} {arXiv:1903.03836 [gr-qc]} \BibitemShut
  {NoStop}%
\bibitem [{\citenamefont {Saridakis}(2020)}]{Saridakis:2020zol}%
  \BibitemOpen
  \bibfield  {author} {\bibinfo {author} {\bibfnamefont {E.~N.}\ \bibnamefont
  {Saridakis}},\ }\href {\doibase 10.1103/PhysRevD.102.123525} {\bibfield
  {journal} {\bibinfo  {journal} {Phys. Rev. D}\ }\textbf {\bibinfo {volume}
  {102}},\ \bibinfo {pages} {123525} (\bibinfo {year} {2020})},\ \Eprint
  {http://arxiv.org/abs/2005.04115} {arXiv:2005.04115 [gr-qc]} \BibitemShut
  {NoStop}%
\bibitem [{\citenamefont {Clifton}\ \emph {et~al.}(2012)\citenamefont
  {Clifton}, \citenamefont {Ferreira}, \citenamefont {Padilla},\ and\
  \citenamefont {Skordis}}]{Clifton:2011jh}%
  \BibitemOpen
  \bibfield  {author} {\bibinfo {author} {\bibfnamefont {T.}~\bibnamefont
  {Clifton}}, \bibinfo {author} {\bibfnamefont {P.~G.}\ \bibnamefont
  {Ferreira}}, \bibinfo {author} {\bibfnamefont {A.}~\bibnamefont {Padilla}}, \
  and\ \bibinfo {author} {\bibfnamefont {C.}~\bibnamefont {Skordis}},\ }\href
  {\doibase 10.1016/j.physrep.2012.01.001} {\bibfield  {journal} {\bibinfo
  {journal} {Phys. Rept.}\ }\textbf {\bibinfo {volume} {513}},\ \bibinfo
  {pages} {1} (\bibinfo {year} {2012})},\ \Eprint
  {http://arxiv.org/abs/1106.2476} {arXiv:1106.2476 [astro-ph.CO]} \BibitemShut
  {NoStop}%
\bibitem [{\citenamefont {Capozziello}\ \emph {et~al.}(2019)\citenamefont
  {Capozziello}, \citenamefont {D'Agostino},\ and\ \citenamefont
  {Luongo}}]{Capozziello:2019cav}%
  \BibitemOpen
  \bibfield  {author} {\bibinfo {author} {\bibfnamefont {S.}~\bibnamefont
  {Capozziello}}, \bibinfo {author} {\bibfnamefont {R.}~\bibnamefont
  {D'Agostino}}, \ and\ \bibinfo {author} {\bibfnamefont {O.}~\bibnamefont
  {Luongo}},\ }\href {\doibase 10.1142/S0218271819300167} {\bibfield  {journal}
  {\bibinfo  {journal} {Int. J. Mod. Phys. D}\ }\textbf {\bibinfo {volume}
  {28}},\ \bibinfo {pages} {1930016} (\bibinfo {year} {2019})},\ \Eprint
  {http://arxiv.org/abs/1904.01427} {arXiv:1904.01427 [gr-qc]} \BibitemShut
  {NoStop}%
\bibitem [{\citenamefont {Starobinsky}(1980)}]{Starobinsky:1980te}%
  \BibitemOpen
  \bibfield  {author} {\bibinfo {author} {\bibfnamefont {A.~A.}\ \bibnamefont
  {Starobinsky}},\ }\href {\doibase 10.1016/0370-2693(80)90670-X} {\bibfield
  {journal} {\bibinfo  {journal} {Phys. Lett. B}\ }\textbf {\bibinfo {volume}
  {91}},\ \bibinfo {pages} {99} (\bibinfo {year} {1980})}\BibitemShut {NoStop}%
\bibitem [{\citenamefont {Guth}(1981)}]{Guth:1980zm}%
  \BibitemOpen
  \bibfield  {author} {\bibinfo {author} {\bibfnamefont {A.~H.}\ \bibnamefont
  {Guth}},\ }\href {\doibase 10.1103/PhysRevD.23.347} {\bibfield  {journal}
  {\bibinfo  {journal} {Phys. Rev. D}\ }\textbf {\bibinfo {volume} {23}},\
  \bibinfo {pages} {347} (\bibinfo {year} {1981})}\BibitemShut {NoStop}%
\bibitem [{\citenamefont {Linde}(1982)}]{Linde:1981mu}%
  \BibitemOpen
  \bibfield  {author} {\bibinfo {author} {\bibfnamefont {A.~D.}\ \bibnamefont
  {Linde}},\ }\href {\doibase 10.1016/0370-2693(82)91219-9} {\bibfield
  {journal} {\bibinfo  {journal} {Phys. Lett. B}\ }\textbf {\bibinfo {volume}
  {108}},\ \bibinfo {pages} {389} (\bibinfo {year} {1982})}\BibitemShut
  {NoStop}%
\bibitem [{\citenamefont {D'Agostino}\ and\ \citenamefont
  {Luongo}(2022)}]{DAgostino:2021vvv}%
  \BibitemOpen
  \bibfield  {author} {\bibinfo {author} {\bibfnamefont {R.}~\bibnamefont
  {D'Agostino}}\ and\ \bibinfo {author} {\bibfnamefont {O.}~\bibnamefont
  {Luongo}},\ }\href {\doibase 10.1016/j.physletb.2022.137070} {\bibfield
  {journal} {\bibinfo  {journal} {Phys. Lett. B}\ }\textbf {\bibinfo {volume}
  {829}},\ \bibinfo {pages} {137070} (\bibinfo {year} {2022})},\ \Eprint
  {http://arxiv.org/abs/2112.12816} {arXiv:2112.12816 [astro-ph.CO]}
  \BibitemShut {NoStop}%
\bibitem [{\citenamefont {Mukhanov}\ \emph {et~al.}(1992)\citenamefont
  {Mukhanov}, \citenamefont {Feldman},\ and\ \citenamefont
  {Brandenberger}}]{Mukhanov:1990me}%
  \BibitemOpen
  \bibfield  {author} {\bibinfo {author} {\bibfnamefont {V.~F.}\ \bibnamefont
  {Mukhanov}}, \bibinfo {author} {\bibfnamefont {H.~A.}\ \bibnamefont
  {Feldman}}, \ and\ \bibinfo {author} {\bibfnamefont {R.~H.}\ \bibnamefont
  {Brandenberger}},\ }\href {\doibase 10.1016/0370-1573(92)90044-Z} {\bibfield
  {journal} {\bibinfo  {journal} {Phys. Rept.}\ }\textbf {\bibinfo {volume}
  {215}},\ \bibinfo {pages} {203} (\bibinfo {year} {1992})}\BibitemShut
  {NoStop}%
\bibitem [{\citenamefont {Di~Valentino}\ \emph {et~al.}(2019)\citenamefont
  {Di~Valentino}, \citenamefont {Melchiorri},\ and\ \citenamefont
  {Silk}}]{DiValentino:2019qzk}%
  \BibitemOpen
  \bibfield  {author} {\bibinfo {author} {\bibfnamefont {E.}~\bibnamefont
  {Di~Valentino}}, \bibinfo {author} {\bibfnamefont {A.}~\bibnamefont
  {Melchiorri}}, \ and\ \bibinfo {author} {\bibfnamefont {J.}~\bibnamefont
  {Silk}},\ }\href {\doibase 10.1038/s41550-019-0906-9} {\bibfield  {journal}
  {\bibinfo  {journal} {Nature Astron.}\ }\textbf {\bibinfo {volume} {4}},\
  \bibinfo {pages} {196} (\bibinfo {year} {2019})},\ \Eprint
  {http://arxiv.org/abs/1911.02087} {arXiv:1911.02087 [astro-ph.CO]}
  \BibitemShut {NoStop}%
\bibitem [{\citenamefont {Handley}(2021)}]{Handley:2019tkm}%
  \BibitemOpen
  \bibfield  {author} {\bibinfo {author} {\bibfnamefont {W.}~\bibnamefont
  {Handley}},\ }\href {\doibase 10.1103/PhysRevD.103.L041301} {\bibfield
  {journal} {\bibinfo  {journal} {Phys. Rev. D}\ }\textbf {\bibinfo {volume}
  {103}},\ \bibinfo {pages} {L041301} (\bibinfo {year} {2021})},\ \Eprint
  {http://arxiv.org/abs/1908.09139} {arXiv:1908.09139 [astro-ph.CO]}
  \BibitemShut {NoStop}%
\bibitem [{\citenamefont {Raveri}\ and\ \citenamefont
  {Hu}(2019)}]{Raveri:2018wln}%
  \BibitemOpen
  \bibfield  {author} {\bibinfo {author} {\bibfnamefont {M.}~\bibnamefont
  {Raveri}}\ and\ \bibinfo {author} {\bibfnamefont {W.}~\bibnamefont {Hu}},\
  }\href {\doibase 10.1103/PhysRevD.99.043506} {\bibfield  {journal} {\bibinfo
  {journal} {Phys. Rev. D}\ }\textbf {\bibinfo {volume} {99}},\ \bibinfo
  {pages} {043506} (\bibinfo {year} {2019})},\ \Eprint
  {http://arxiv.org/abs/1806.04649} {arXiv:1806.04649 [astro-ph.CO]}
  \BibitemShut {NoStop}%
\bibitem [{\citenamefont {O'Dwyer}\ \emph {et~al.}(2020)\citenamefont
  {O'Dwyer}, \citenamefont {Anselmi}, \citenamefont {Starkman}, \citenamefont
  {Corasaniti}, \citenamefont {Sheth},\ and\ \citenamefont
  {Zehavi}}]{ODwyer:2019rvi}%
  \BibitemOpen
  \bibfield  {author} {\bibinfo {author} {\bibfnamefont {M.}~\bibnamefont
  {O'Dwyer}}, \bibinfo {author} {\bibfnamefont {S.}~\bibnamefont {Anselmi}},
  \bibinfo {author} {\bibfnamefont {G.~D.}\ \bibnamefont {Starkman}}, \bibinfo
  {author} {\bibfnamefont {P.-S.}\ \bibnamefont {Corasaniti}}, \bibinfo
  {author} {\bibfnamefont {R.~K.}\ \bibnamefont {Sheth}}, \ and\ \bibinfo
  {author} {\bibfnamefont {I.}~\bibnamefont {Zehavi}},\ }\href {\doibase
  10.1103/PhysRevD.101.083517} {\bibfield  {journal} {\bibinfo  {journal}
  {Phys. Rev. D}\ }\textbf {\bibinfo {volume} {101}},\ \bibinfo {pages}
  {083517} (\bibinfo {year} {2020})},\ \Eprint
  {http://arxiv.org/abs/1910.10698} {arXiv:1910.10698 [astro-ph.CO]}
  \BibitemShut {NoStop}%
\bibitem [{\citenamefont {Glanville}\ \emph {et~al.}(2022)\citenamefont
  {Glanville}, \citenamefont {Howlett},\ and\ \citenamefont
  {Davis}}]{Glanville:2022xes}%
  \BibitemOpen
  \bibfield  {author} {\bibinfo {author} {\bibfnamefont {A.}~\bibnamefont
  {Glanville}}, \bibinfo {author} {\bibfnamefont {C.}~\bibnamefont {Howlett}},
  \ and\ \bibinfo {author} {\bibfnamefont {T.~M.}\ \bibnamefont {Davis}},\
  }\href {\doibase 10.1093/mnras/stac2891} {\bibfield  {journal} {\bibinfo
  {journal} {Mon. Not. Roy. Astron. Soc.}\ }\textbf {\bibinfo {volume} {517}},\
  \bibinfo {pages} {3087} (\bibinfo {year} {2022})},\ \Eprint
  {http://arxiv.org/abs/2205.05892} {arXiv:2205.05892 [astro-ph.CO]}
  \BibitemShut {NoStop}%
\bibitem [{\citenamefont {Lewis}\ \emph {et~al.}(2000)\citenamefont {Lewis},
  \citenamefont {Challinor},\ and\ \citenamefont {Lasenby}}]{Lewis:1999bs}%
  \BibitemOpen
  \bibfield  {author} {\bibinfo {author} {\bibfnamefont {A.}~\bibnamefont
  {Lewis}}, \bibinfo {author} {\bibfnamefont {A.}~\bibnamefont {Challinor}}, \
  and\ \bibinfo {author} {\bibfnamefont {A.}~\bibnamefont {Lasenby}},\ }\href
  {\doibase 10.1086/309179} {\bibfield  {journal} {\bibinfo  {journal}
  {Astrophys. J.}\ }\textbf {\bibinfo {volume} {538}},\ \bibinfo {pages} {473}
  (\bibinfo {year} {2000})},\ \Eprint {http://arxiv.org/abs/astro-ph/9911177}
  {arXiv:astro-ph/9911177} \BibitemShut {NoStop}%
\bibitem [{\citenamefont {Masso}\ \emph {et~al.}(2008)\citenamefont {Masso},
  \citenamefont {Mohanty}, \citenamefont {Nautiyal},\ and\ \citenamefont
  {Zsembinszki}}]{Masso:2006gv}%
  \BibitemOpen
  \bibfield  {author} {\bibinfo {author} {\bibfnamefont {E.}~\bibnamefont
  {Masso}}, \bibinfo {author} {\bibfnamefont {S.}~\bibnamefont {Mohanty}},
  \bibinfo {author} {\bibfnamefont {A.}~\bibnamefont {Nautiyal}}, \ and\
  \bibinfo {author} {\bibfnamefont {G.}~\bibnamefont {Zsembinszki}},\ }\href
  {\doibase 10.1103/PhysRevD.78.043534} {\bibfield  {journal} {\bibinfo
  {journal} {Phys. Rev. D}\ }\textbf {\bibinfo {volume} {78}},\ \bibinfo
  {pages} {043534} (\bibinfo {year} {2008})},\ \Eprint
  {http://arxiv.org/abs/astro-ph/0609349} {arXiv:astro-ph/0609349} \BibitemShut
  {NoStop}%
\bibitem [{\citenamefont {Lasenby}\ and\ \citenamefont
  {Doran}(2005)}]{Lasenby:2003ur}%
  \BibitemOpen
  \bibfield  {author} {\bibinfo {author} {\bibfnamefont {A.}~\bibnamefont
  {Lasenby}}\ and\ \bibinfo {author} {\bibfnamefont {C.}~\bibnamefont
  {Doran}},\ }\href {\doibase 10.1103/PhysRevD.71.063502} {\bibfield  {journal}
  {\bibinfo  {journal} {Phys. Rev. D}\ }\textbf {\bibinfo {volume} {71}},\
  \bibinfo {pages} {063502} (\bibinfo {year} {2005})},\ \Eprint
  {http://arxiv.org/abs/astro-ph/0307311} {arXiv:astro-ph/0307311} \BibitemShut
  {NoStop}%
\bibitem [{\citenamefont {Efstathiou}(2003)}]{Efstathiou:2003hk}%
  \BibitemOpen
  \bibfield  {author} {\bibinfo {author} {\bibfnamefont {G.}~\bibnamefont
  {Efstathiou}},\ }\href {\doibase 10.1046/j.1365-8711.2003.06940.x} {\bibfield
   {journal} {\bibinfo  {journal} {Mon. Not. Roy. Astron. Soc.}\ }\textbf
  {\bibinfo {volume} {343}},\ \bibinfo {pages} {L95} (\bibinfo {year}
  {2003})},\ \Eprint {http://arxiv.org/abs/astro-ph/0303127}
  {arXiv:astro-ph/0303127} \BibitemShut {NoStop}%
\bibitem [{\citenamefont {Guzzetti}\ \emph {et~al.}(2016)\citenamefont
  {Guzzetti}, \citenamefont {Bartolo}, \citenamefont {Liguori},\ and\
  \citenamefont {Matarrese}}]{Guzzetti:2016mkm}%
  \BibitemOpen
  \bibfield  {author} {\bibinfo {author} {\bibfnamefont {M.~C.}\ \bibnamefont
  {Guzzetti}}, \bibinfo {author} {\bibfnamefont {N.}~\bibnamefont {Bartolo}},
  \bibinfo {author} {\bibfnamefont {M.}~\bibnamefont {Liguori}}, \ and\
  \bibinfo {author} {\bibfnamefont {S.}~\bibnamefont {Matarrese}},\ }\href
  {\doibase 10.1393/ncr/i2016-10127-1} {\bibfield  {journal} {\bibinfo
  {journal} {Riv. Nuovo Cim.}\ }\textbf {\bibinfo {volume} {39}},\ \bibinfo
  {pages} {399} (\bibinfo {year} {2016})},\ \Eprint
  {http://arxiv.org/abs/1605.01615} {arXiv:1605.01615 [astro-ph.CO]}
  \BibitemShut {NoStop}%
\bibitem [{\citenamefont {Armitage-Caplan}\ \emph {et~al.}(2011)\citenamefont
  {Armitage-Caplan} \emph {et~al.}}]{COrE}%
  \BibitemOpen
  \bibfield  {author} {\bibinfo {author} {\bibfnamefont {C.}~\bibnamefont
  {Armitage-Caplan}} \emph {et~al.} (\bibinfo {collaboration} {COrE
  Collaboration}),\ }\href@noop {} {\  (\bibinfo {year} {2011})},\ \Eprint
  {http://arxiv.org/abs/1102.2181} {arXiv:1102.2181 [astro-ph.CO]} \BibitemShut
  {NoStop}%
\bibitem [{\citenamefont {Andr\'e}\ \emph {et~al.}(2014)\citenamefont {Andr\'e}
  \emph {et~al.}}]{PRISM:2013fvg}%
  \BibitemOpen
  \bibfield  {author} {\bibinfo {author} {\bibfnamefont {P.}~\bibnamefont
  {Andr\'e}} \emph {et~al.} (\bibinfo {collaboration} {PRISM}),\ }\href
  {\doibase 10.1088/1475-7516/2014/02/006} {\bibfield  {journal} {\bibinfo
  {journal} {JCAP}\ }\textbf {\bibinfo {volume} {02}},\ \bibinfo {pages} {006}
  (\bibinfo {year} {2014})},\ \Eprint {http://arxiv.org/abs/1310.1554}
  {arXiv:1310.1554 [astro-ph.CO]} \BibitemShut {NoStop}%
\bibitem [{\citenamefont {Allys}\ \emph {et~al.}(2022)\citenamefont {Allys}
  \emph {et~al.}}]{LiteBIRD:2022cnt}%
  \BibitemOpen
  \bibfield  {author} {\bibinfo {author} {\bibfnamefont {E.}~\bibnamefont
  {Allys}} \emph {et~al.} (\bibinfo {collaboration} {LiteBIRD}),\ }\href
  {\doibase 10.1093/ptep/ptac150} {\  (\bibinfo {year} {2022}),\
  10.1093/ptep/ptac150},\ \Eprint {http://arxiv.org/abs/2202.02773}
  {arXiv:2202.02773 [astro-ph.IM]} \BibitemShut {NoStop}%
\bibitem [{\citenamefont {Dodelson}\ \emph {et~al.}(2003)\citenamefont
  {Dodelson}, \citenamefont {Rozo},\ and\ \citenamefont
  {Stebbins}}]{Dodelson:2003bv}%
  \BibitemOpen
  \bibfield  {author} {\bibinfo {author} {\bibfnamefont {S.}~\bibnamefont
  {Dodelson}}, \bibinfo {author} {\bibfnamefont {E.}~\bibnamefont {Rozo}}, \
  and\ \bibinfo {author} {\bibfnamefont {A.}~\bibnamefont {Stebbins}},\ }\href
  {\doibase 10.1103/PhysRevLett.91.021301} {\bibfield  {journal} {\bibinfo
  {journal} {Phys. Rev. Lett.}\ }\textbf {\bibinfo {volume} {91}},\ \bibinfo
  {pages} {021301} (\bibinfo {year} {2003})},\ \Eprint
  {http://arxiv.org/abs/astro-ph/0301177} {arXiv:astro-ph/0301177} \BibitemShut
  {NoStop}%
\bibitem [{\citenamefont {Masui}\ and\ \citenamefont
  {Pen}(2010)}]{Masui:2010cz}%
  \BibitemOpen
  \bibfield  {author} {\bibinfo {author} {\bibfnamefont {K.~W.}\ \bibnamefont
  {Masui}}\ and\ \bibinfo {author} {\bibfnamefont {U.-L.}\ \bibnamefont
  {Pen}},\ }\href {\doibase 10.1103/PhysRevLett.105.161302} {\bibfield
  {journal} {\bibinfo  {journal} {Phys. Rev. Lett.}\ }\textbf {\bibinfo
  {volume} {105}},\ \bibinfo {pages} {161302} (\bibinfo {year} {2010})},\
  \Eprint {http://arxiv.org/abs/1006.4181} {arXiv:1006.4181 [astro-ph.CO]}
  \BibitemShut {NoStop}%
\bibitem [{\citenamefont {Smith}\ \emph {et~al.}(2006)\citenamefont {Smith},
  \citenamefont {Pierpaoli},\ and\ \citenamefont
  {Kamionkowski}}]{Smith:2006nka}%
  \BibitemOpen
  \bibfield  {author} {\bibinfo {author} {\bibfnamefont {T.~L.}\ \bibnamefont
  {Smith}}, \bibinfo {author} {\bibfnamefont {E.}~\bibnamefont {Pierpaoli}}, \
  and\ \bibinfo {author} {\bibfnamefont {M.}~\bibnamefont {Kamionkowski}},\
  }\href {\doibase 10.1103/PhysRevLett.97.021301} {\bibfield  {journal}
  {\bibinfo  {journal} {Phys. Rev. Lett.}\ }\textbf {\bibinfo {volume} {97}},\
  \bibinfo {pages} {021301} (\bibinfo {year} {2006})},\ \Eprint
  {http://arxiv.org/abs/astro-ph/0603144} {arXiv:astro-ph/0603144} \BibitemShut
  {NoStop}%
\bibitem [{\citenamefont {Maggiore}\ \emph {et~al.}(2020)\citenamefont
  {Maggiore} \emph {et~al.}}]{Maggiore:2019uih}%
  \BibitemOpen
  \bibfield  {author} {\bibinfo {author} {\bibfnamefont {M.}~\bibnamefont
  {Maggiore}} \emph {et~al.},\ }\href {\doibase 10.1088/1475-7516/2020/03/050}
  {\bibfield  {journal} {\bibinfo  {journal} {JCAP}\ }\textbf {\bibinfo
  {volume} {03}},\ \bibinfo {pages} {050} (\bibinfo {year} {2020})},\ \Eprint
  {http://arxiv.org/abs/1912.02622} {arXiv:1912.02622 [astro-ph.CO]}
  \BibitemShut {NoStop}%
\bibitem [{\citenamefont {Branchesi}\ \emph {et~al.}(2023)\citenamefont
  {Branchesi} \emph {et~al.}}]{Branchesi:2023mws}%
  \BibitemOpen
  \bibfield  {author} {\bibinfo {author} {\bibfnamefont {M.}~\bibnamefont
  {Branchesi}} \emph {et~al.},\ }\href@noop {} {\  (\bibinfo {year} {2023})},\
  \Eprint {http://arxiv.org/abs/2303.15923} {arXiv:2303.15923 [gr-qc]}
  \BibitemShut {NoStop}%
\bibitem [{\citenamefont {Amaro-Seoane}\ \emph {et~al.}(2017)\citenamefont
  {Amaro-Seoane} \emph {et~al.}}]{LISA:2017pwj}%
  \BibitemOpen
  \bibfield  {author} {\bibinfo {author} {\bibfnamefont {P.}~\bibnamefont
  {Amaro-Seoane}} \emph {et~al.} (\bibinfo {collaboration} {LISA}),\
  }\href@noop {} {\  (\bibinfo {year} {2017})},\ \Eprint
  {http://arxiv.org/abs/1702.00786} {arXiv:1702.00786 [astro-ph.IM]}
  \BibitemShut {NoStop}%
\bibitem [{\citenamefont {Kawamura}\ \emph {et~al.}(2021)\citenamefont
  {Kawamura} \emph {et~al.}}]{Kawamura:2020pcg}%
  \BibitemOpen
  \bibfield  {author} {\bibinfo {author} {\bibfnamefont {S.}~\bibnamefont
  {Kawamura}} \emph {et~al.},\ }\href {\doibase 10.1093/ptep/ptab019}
  {\bibfield  {journal} {\bibinfo  {journal} {PTEP}\ }\textbf {\bibinfo
  {volume} {2021}},\ \bibinfo {pages} {05A105} (\bibinfo {year} {2021})},\
  \Eprint {http://arxiv.org/abs/2006.13545} {arXiv:2006.13545 [gr-qc]}
  \BibitemShut {NoStop}%
\bibitem [{\citenamefont {Cai}\ \emph {et~al.}(2018)\citenamefont {Cai},
  \citenamefont {Liu}, \citenamefont {Liu}, \citenamefont {Wang},\ and\
  \citenamefont {Yang}}]{Cai:2017aea}%
  \BibitemOpen
  \bibfield  {author} {\bibinfo {author} {\bibfnamefont {R.-G.}\ \bibnamefont
  {Cai}}, \bibinfo {author} {\bibfnamefont {T.-B.}\ \bibnamefont {Liu}},
  \bibinfo {author} {\bibfnamefont {X.-W.}\ \bibnamefont {Liu}}, \bibinfo
  {author} {\bibfnamefont {S.-J.}\ \bibnamefont {Wang}}, \ and\ \bibinfo
  {author} {\bibfnamefont {T.}~\bibnamefont {Yang}},\ }\href {\doibase
  10.1103/PhysRevD.97.103005} {\bibfield  {journal} {\bibinfo  {journal} {Phys.
  Rev. D}\ }\textbf {\bibinfo {volume} {97}},\ \bibinfo {pages} {103005}
  (\bibinfo {year} {2018})},\ \Eprint {http://arxiv.org/abs/1712.00952}
  {arXiv:1712.00952 [astro-ph.CO]} \BibitemShut {NoStop}%
\bibitem [{\citenamefont {Belgacem}\ \emph {et~al.}(2018)\citenamefont
  {Belgacem}, \citenamefont {Dirian}, \citenamefont {Foffa},\ and\
  \citenamefont {Maggiore}}]{Belgacem:2017ihm}%
  \BibitemOpen
  \bibfield  {author} {\bibinfo {author} {\bibfnamefont {E.}~\bibnamefont
  {Belgacem}}, \bibinfo {author} {\bibfnamefont {Y.}~\bibnamefont {Dirian}},
  \bibinfo {author} {\bibfnamefont {S.}~\bibnamefont {Foffa}}, \ and\ \bibinfo
  {author} {\bibfnamefont {M.}~\bibnamefont {Maggiore}},\ }\href {\doibase
  10.1103/PhysRevD.97.104066} {\bibfield  {journal} {\bibinfo  {journal} {Phys.
  Rev. D}\ }\textbf {\bibinfo {volume} {97}},\ \bibinfo {pages} {104066}
  (\bibinfo {year} {2018})},\ \Eprint {http://arxiv.org/abs/1712.08108}
  {arXiv:1712.08108 [astro-ph.CO]} \BibitemShut {NoStop}%
\bibitem [{\citenamefont {D'Agostino}\ and\ \citenamefont
  {Nunes}(2019)}]{DAgostino:2019hvh}%
  \BibitemOpen
  \bibfield  {author} {\bibinfo {author} {\bibfnamefont {R.}~\bibnamefont
  {D'Agostino}}\ and\ \bibinfo {author} {\bibfnamefont {R.~C.}\ \bibnamefont
  {Nunes}},\ }\href {\doibase 10.1103/PhysRevD.100.044041} {\bibfield
  {journal} {\bibinfo  {journal} {Phys. Rev. D}\ }\textbf {\bibinfo {volume}
  {100}},\ \bibinfo {pages} {044041} (\bibinfo {year} {2019})},\ \Eprint
  {http://arxiv.org/abs/1907.05516} {arXiv:1907.05516 [gr-qc]} \BibitemShut
  {NoStop}%
\bibitem [{\citenamefont {Bonilla}\ \emph {et~al.}(2020)\citenamefont
  {Bonilla}, \citenamefont {D'Agostino}, \citenamefont {Nunes},\ and\
  \citenamefont {de~Araujo}}]{Bonilla:2019mbm}%
  \BibitemOpen
  \bibfield  {author} {\bibinfo {author} {\bibfnamefont {A.}~\bibnamefont
  {Bonilla}}, \bibinfo {author} {\bibfnamefont {R.}~\bibnamefont {D'Agostino}},
  \bibinfo {author} {\bibfnamefont {R.~C.}\ \bibnamefont {Nunes}}, \ and\
  \bibinfo {author} {\bibfnamefont {J.~C.~N.}\ \bibnamefont {de~Araujo}},\
  }\href {\doibase 10.1088/1475-7516/2020/03/015} {\bibfield  {journal}
  {\bibinfo  {journal} {JCAP}\ }\textbf {\bibinfo {volume} {03}},\ \bibinfo
  {pages} {015} (\bibinfo {year} {2020})},\ \Eprint
  {http://arxiv.org/abs/1910.05631} {arXiv:1910.05631 [gr-qc]} \BibitemShut
  {NoStop}%
\bibitem [{\citenamefont {Baker}\ and\ \citenamefont
  {Harrison}(2021)}]{Baker:2020apq}%
  \BibitemOpen
  \bibfield  {author} {\bibinfo {author} {\bibfnamefont {T.}~\bibnamefont
  {Baker}}\ and\ \bibinfo {author} {\bibfnamefont {I.}~\bibnamefont
  {Harrison}},\ }\href {\doibase 10.1088/1475-7516/2021/01/068} {\bibfield
  {journal} {\bibinfo  {journal} {JCAP}\ }\textbf {\bibinfo {volume} {01}},\
  \bibinfo {pages} {068} (\bibinfo {year} {2021})},\ \Eprint
  {http://arxiv.org/abs/2007.13791} {arXiv:2007.13791 [astro-ph.CO]}
  \BibitemShut {NoStop}%
\bibitem [{\citenamefont {Tasinato}\ \emph {et~al.}(2021)\citenamefont
  {Tasinato}, \citenamefont {Garoffolo}, \citenamefont {Bertacca},\ and\
  \citenamefont {Matarrese}}]{Tasinato:2021wol}%
  \BibitemOpen
  \bibfield  {author} {\bibinfo {author} {\bibfnamefont {G.}~\bibnamefont
  {Tasinato}}, \bibinfo {author} {\bibfnamefont {A.}~\bibnamefont {Garoffolo}},
  \bibinfo {author} {\bibfnamefont {D.}~\bibnamefont {Bertacca}}, \ and\
  \bibinfo {author} {\bibfnamefont {S.}~\bibnamefont {Matarrese}},\ }\href
  {\doibase 10.1088/1475-7516/2021/06/050} {\bibfield  {journal} {\bibinfo
  {journal} {JCAP}\ }\textbf {\bibinfo {volume} {06}},\ \bibinfo {pages} {050}
  (\bibinfo {year} {2021})},\ \Eprint {http://arxiv.org/abs/2103.00155}
  {arXiv:2103.00155 [gr-qc]} \BibitemShut {NoStop}%
\bibitem [{\citenamefont {Ezquiaga}\ \emph {et~al.}(2021)\citenamefont
  {Ezquiaga}, \citenamefont {Hu}, \citenamefont {Lagos},\ and\ \citenamefont
  {Lin}}]{Ezquiaga:2021ler}%
  \BibitemOpen
  \bibfield  {author} {\bibinfo {author} {\bibfnamefont {J.~M.}\ \bibnamefont
  {Ezquiaga}}, \bibinfo {author} {\bibfnamefont {W.}~\bibnamefont {Hu}},
  \bibinfo {author} {\bibfnamefont {M.}~\bibnamefont {Lagos}}, \ and\ \bibinfo
  {author} {\bibfnamefont {M.-X.}\ \bibnamefont {Lin}},\ }\href {\doibase
  10.1088/1475-7516/2021/11/048} {\bibfield  {journal} {\bibinfo  {journal}
  {JCAP}\ }\textbf {\bibinfo {volume} {11}},\ \bibinfo {pages} {048} (\bibinfo
  {year} {2021})},\ \Eprint {http://arxiv.org/abs/2108.10872} {arXiv:2108.10872
  [astro-ph.CO]} \BibitemShut {NoStop}%
\bibitem [{\citenamefont {Califano}\ \emph {et~al.}(2022)\citenamefont
  {Califano}, \citenamefont {de~Martino}, \citenamefont {Vernieri},\ and\
  \citenamefont {Capozziello}}]{Califano:2022cmo}%
  \BibitemOpen
  \bibfield  {author} {\bibinfo {author} {\bibfnamefont {M.}~\bibnamefont
  {Califano}}, \bibinfo {author} {\bibfnamefont {I.}~\bibnamefont
  {de~Martino}}, \bibinfo {author} {\bibfnamefont {D.}~\bibnamefont
  {Vernieri}}, \ and\ \bibinfo {author} {\bibfnamefont {S.}~\bibnamefont
  {Capozziello}},\ }\href {\doibase 10.1093/mnras/stac3230} {\  (\bibinfo
  {year} {2022}),\ 10.1093/mnras/stac3230},\ \Eprint
  {http://arxiv.org/abs/2205.11221} {arXiv:2205.11221 [astro-ph.CO]}
  \BibitemShut {NoStop}%
\bibitem [{\citenamefont {Badger}\ and\ \citenamefont
  {Sakellariadou}(2021)}]{Badger:2021enh}%
  \BibitemOpen
  \bibfield  {author} {\bibinfo {author} {\bibfnamefont {C.}~\bibnamefont
  {Badger}}\ and\ \bibinfo {author} {\bibfnamefont {M.}~\bibnamefont
  {Sakellariadou}},\ }\href@noop {} {\  (\bibinfo {year} {2021})},\ \Eprint
  {http://arxiv.org/abs/2112.04650} {arXiv:2112.04650 [gr-qc]} \BibitemShut
  {NoStop}%
\bibitem [{\citenamefont {D'Agostino}\ and\ \citenamefont
  {Nunes}(2022)}]{DAgostino:2022tdk}%
  \BibitemOpen
  \bibfield  {author} {\bibinfo {author} {\bibfnamefont {R.}~\bibnamefont
  {D'Agostino}}\ and\ \bibinfo {author} {\bibfnamefont {R.~C.}\ \bibnamefont
  {Nunes}},\ }\href {\doibase 10.1103/PhysRevD.106.124053} {\bibfield
  {journal} {\bibinfo  {journal} {Phys. Rev. D}\ }\textbf {\bibinfo {volume}
  {106}},\ \bibinfo {pages} {124053} (\bibinfo {year} {2022})},\ \Eprint
  {http://arxiv.org/abs/2210.11935} {arXiv:2210.11935 [gr-qc]} \BibitemShut
  {NoStop}%
\bibitem [{\citenamefont {Califano}\ \emph {et~al.}(2023)\citenamefont
  {Califano}, \citenamefont {de~Martino}, \citenamefont {Vernieri},\ and\
  \citenamefont {Capozziello}}]{Califano:2022syd}%
  \BibitemOpen
  \bibfield  {author} {\bibinfo {author} {\bibfnamefont {M.}~\bibnamefont
  {Califano}}, \bibinfo {author} {\bibfnamefont {I.}~\bibnamefont
  {de~Martino}}, \bibinfo {author} {\bibfnamefont {D.}~\bibnamefont
  {Vernieri}}, \ and\ \bibinfo {author} {\bibfnamefont {S.}~\bibnamefont
  {Capozziello}},\ }\href {\doibase 10.1103/PhysRevD.107.123519} {\bibfield
  {journal} {\bibinfo  {journal} {Phys. Rev. D}\ }\textbf {\bibinfo {volume}
  {107}},\ \bibinfo {pages} {123519} (\bibinfo {year} {2023})},\ \Eprint
  {http://arxiv.org/abs/2208.13999} {arXiv:2208.13999 [astro-ph.CO]}
  \BibitemShut {NoStop}%
\bibitem [{\citenamefont {Pieroni}\ \emph {et~al.}(2022)\citenamefont
  {Pieroni}, \citenamefont {Ricciardone},\ and\ \citenamefont
  {Barausse}}]{Pieroni:2022bbh}%
  \BibitemOpen
  \bibfield  {author} {\bibinfo {author} {\bibfnamefont {M.}~\bibnamefont
  {Pieroni}}, \bibinfo {author} {\bibfnamefont {A.}~\bibnamefont
  {Ricciardone}}, \ and\ \bibinfo {author} {\bibfnamefont {E.}~\bibnamefont
  {Barausse}},\ }\href {\doibase 10.1038/s41598-022-19540-7} {\bibfield
  {journal} {\bibinfo  {journal} {Sci. Rep.}\ }\textbf {\bibinfo {volume}
  {12}},\ \bibinfo {pages} {17940} (\bibinfo {year} {2022})},\ \Eprint
  {http://arxiv.org/abs/2203.12586} {arXiv:2203.12586 [astro-ph.CO]}
  \BibitemShut {NoStop}%
\bibitem [{\citenamefont {Babak}\ \emph {et~al.}(2016)\citenamefont {Babak}
  \emph {et~al.}}]{Babak:2015lua}%
  \BibitemOpen
  \bibfield  {author} {\bibinfo {author} {\bibfnamefont {S.}~\bibnamefont
  {Babak}} \emph {et~al.},\ }\href {\doibase 10.1093/mnras/stv2092} {\bibfield
  {journal} {\bibinfo  {journal} {Mon. Not. Roy. Astron. Soc.}\ }\textbf
  {\bibinfo {volume} {455}},\ \bibinfo {pages} {1665} (\bibinfo {year}
  {2016})},\ \Eprint {http://arxiv.org/abs/1509.02165} {arXiv:1509.02165
  [astro-ph.CO]} \BibitemShut {NoStop}%
\bibitem [{\citenamefont {Desvignes}\ \emph {et~al.}(2016)\citenamefont
  {Desvignes} \emph {et~al.}}]{Desvignes:2016yex}%
  \BibitemOpen
  \bibfield  {author} {\bibinfo {author} {\bibfnamefont {G.}~\bibnamefont
  {Desvignes}} \emph {et~al.},\ }\href {\doibase 10.1093/mnras/stw483}
  {\bibfield  {journal} {\bibinfo  {journal} {Mon. Not. Roy. Astron. Soc.}\
  }\textbf {\bibinfo {volume} {458}},\ \bibinfo {pages} {3341} (\bibinfo {year}
  {2016})},\ \Eprint {http://arxiv.org/abs/1602.08511} {arXiv:1602.08511
  [astro-ph.HE]} \BibitemShut {NoStop}%
\bibitem [{\citenamefont {Arzoumanian}\ \emph {et~al.}(2018)\citenamefont
  {Arzoumanian} \emph {et~al.}}]{NANOGRAV:2018hou}%
  \BibitemOpen
  \bibfield  {author} {\bibinfo {author} {\bibfnamefont {Z.}~\bibnamefont
  {Arzoumanian}} \emph {et~al.} (\bibinfo {collaboration} {NANOGRAV}),\ }\href
  {\doibase 10.3847/1538-4357/aabd3b} {\bibfield  {journal} {\bibinfo
  {journal} {Astrophys. J.}\ }\textbf {\bibinfo {volume} {859}},\ \bibinfo
  {pages} {47} (\bibinfo {year} {2018})},\ \Eprint
  {http://arxiv.org/abs/1801.02617} {arXiv:1801.02617 [astro-ph.HE]}
  \BibitemShut {NoStop}%
\bibitem [{\citenamefont {Lasky}\ \emph {et~al.}(2016)\citenamefont {Lasky},
  \citenamefont {Mingarelli}, \citenamefont {Smith}, \citenamefont {Giblin},
  \citenamefont {Thrane}, \citenamefont {Reardon} \emph {et~al.}}]{PPTA}%
  \BibitemOpen
  \bibfield  {author} {\bibinfo {author} {\bibfnamefont {P.~D.}\ \bibnamefont
  {Lasky}}, \bibinfo {author} {\bibfnamefont {C.~M.~F.}\ \bibnamefont
  {Mingarelli}}, \bibinfo {author} {\bibfnamefont {T.~L.}\ \bibnamefont
  {Smith}}, \bibinfo {author} {\bibfnamefont {J.~T.}\ \bibnamefont {Giblin}},
  \bibinfo {author} {\bibfnamefont {E.}~\bibnamefont {Thrane}}, \bibinfo
  {author} {\bibfnamefont {D.~J.}\ \bibnamefont {Reardon}},  \emph {et~al.},\
  }\href {\doibase 10.1103/PhysRevX.6.011035} {\bibfield  {journal} {\bibinfo
  {journal} {Phys. Rev. X}\ }\textbf {\bibinfo {volume} {6}},\ \bibinfo {pages}
  {011035} (\bibinfo {year} {2016})}\BibitemShut {NoStop}%
\bibitem [{\citenamefont {{Verbiest}}\ \emph {et~al.}(2016)\citenamefont
  {{Verbiest}} \emph {et~al.}}]{IPTA}%
  \BibitemOpen
  \bibfield  {author} {\bibinfo {author} {\bibfnamefont {J.~P.~W.}\
  \bibnamefont {{Verbiest}}} \emph {et~al.},\ }\href {\doibase
  10.1093/mnras/stw347} {\bibfield  {journal} {\bibinfo  {journal} {Mon. Not.
  Roy. Astron. Soc.}\ }\textbf {\bibinfo {volume} {458}},\ \bibinfo {pages}
  {1267} (\bibinfo {year} {2016})},\ \Eprint {http://arxiv.org/abs/1602.03640}
  {arXiv:1602.03640 [astro-ph.IM]} \BibitemShut {NoStop}%
\bibitem [{\citenamefont {Abbott}\ and\ \citenamefont
  {Schaefer}(1986)}]{Abbott:1986ct}%
  \BibitemOpen
  \bibfield  {author} {\bibinfo {author} {\bibfnamefont {L.~F.}\ \bibnamefont
  {Abbott}}\ and\ \bibinfo {author} {\bibfnamefont {R.~K.}\ \bibnamefont
  {Schaefer}},\ }\href {\doibase 10.1086/164525} {\bibfield  {journal}
  {\bibinfo  {journal} {Astrophys. J.}\ }\textbf {\bibinfo {volume} {308}},\
  \bibinfo {pages} {546} (\bibinfo {year} {1986})}\BibitemShut {NoStop}%
\bibitem [{\citenamefont {Hu}\ \emph {et~al.}(1998)\citenamefont {Hu},
  \citenamefont {Seljak}, \citenamefont {White},\ and\ \citenamefont
  {Zaldarriaga}}]{Hu:1997mn}%
  \BibitemOpen
  \bibfield  {author} {\bibinfo {author} {\bibfnamefont {W.}~\bibnamefont
  {Hu}}, \bibinfo {author} {\bibfnamefont {U.}~\bibnamefont {Seljak}}, \bibinfo
  {author} {\bibfnamefont {M.~J.}\ \bibnamefont {White}}, \ and\ \bibinfo
  {author} {\bibfnamefont {M.}~\bibnamefont {Zaldarriaga}},\ }\href {\doibase
  10.1103/PhysRevD.57.3290} {\bibfield  {journal} {\bibinfo  {journal} {Phys.
  Rev. D}\ }\textbf {\bibinfo {volume} {57}},\ \bibinfo {pages} {3290}
  (\bibinfo {year} {1998})},\ \Eprint {http://arxiv.org/abs/astro-ph/9709066}
  {arXiv:astro-ph/9709066} \BibitemShut {NoStop}%
\bibitem [{\citenamefont {Akama}\ and\ \citenamefont
  {Kobayashi}(2019)}]{Akama:2018cqv}%
  \BibitemOpen
  \bibfield  {author} {\bibinfo {author} {\bibfnamefont {S.}~\bibnamefont
  {Akama}}\ and\ \bibinfo {author} {\bibfnamefont {T.}~\bibnamefont
  {Kobayashi}},\ }\href {\doibase 10.1103/PhysRevD.99.043522} {\bibfield
  {journal} {\bibinfo  {journal} {Phys. Rev. D}\ }\textbf {\bibinfo {volume}
  {99}},\ \bibinfo {pages} {043522} (\bibinfo {year} {2019})},\ \Eprint
  {http://arxiv.org/abs/1810.01863} {arXiv:1810.01863 [gr-qc]} \BibitemShut
  {NoStop}%
\bibitem [{\citenamefont {Watanabe}\ and\ \citenamefont
  {Komatsu}(2006)}]{Watanabe:2006qe}%
  \BibitemOpen
  \bibfield  {author} {\bibinfo {author} {\bibfnamefont {Y.}~\bibnamefont
  {Watanabe}}\ and\ \bibinfo {author} {\bibfnamefont {E.}~\bibnamefont
  {Komatsu}},\ }\href {\doibase 10.1103/PhysRevD.73.123515} {\bibfield
  {journal} {\bibinfo  {journal} {Phys. Rev. D}\ }\textbf {\bibinfo {volume}
  {73}},\ \bibinfo {pages} {123515} (\bibinfo {year} {2006})},\ \Eprint
  {http://arxiv.org/abs/astro-ph/0604176} {arXiv:astro-ph/0604176} \BibitemShut
  {NoStop}%
\bibitem [{\citenamefont {Baumann}(2018)}]{Baumann:2018muz}%
  \BibitemOpen
  \bibfield  {author} {\bibinfo {author} {\bibfnamefont {D.}~\bibnamefont
  {Baumann}},\ }\href {\doibase 10.22323/1.305.0009} {\bibfield  {journal}
  {\bibinfo  {journal} {PoS}\ }\textbf {\bibinfo {volume} {TASI2017}},\
  \bibinfo {pages} {009} (\bibinfo {year} {2018})},\ \Eprint
  {http://arxiv.org/abs/1807.03098} {arXiv:1807.03098 [hep-th]} \BibitemShut
  {NoStop}%
\bibitem [{\citenamefont {Bunch}\ and\ \citenamefont
  {Davies}(1978)}]{Bunch:1978yq}%
  \BibitemOpen
  \bibfield  {author} {\bibinfo {author} {\bibfnamefont {T.~S.}\ \bibnamefont
  {Bunch}}\ and\ \bibinfo {author} {\bibfnamefont {P.~C.~W.}\ \bibnamefont
  {Davies}},\ }\href {\doibase 10.1098/rspa.1978.0060} {\bibfield  {journal}
  {\bibinfo  {journal} {Proc. Roy. Soc. Lond. A}\ }\textbf {\bibinfo {volume}
  {360}},\ \bibinfo {pages} {117} (\bibinfo {year} {1978})}\BibitemShut
  {NoStop}%
\bibitem [{\citenamefont {Birrell}\ and\ \citenamefont
  {Davies}(1984)}]{Birrell:1982ix}%
  \BibitemOpen
  \bibfield  {author} {\bibinfo {author} {\bibfnamefont {N.~D.}\ \bibnamefont
  {Birrell}}\ and\ \bibinfo {author} {\bibfnamefont {P.~C.~W.}\ \bibnamefont
  {Davies}},\ }\href {\doibase 10.1017/CBO9780511622632} {\emph {\bibinfo
  {title} {{Quantum Fields in Curved Space}}}},\ Cambridge Monographs on
  Mathematical Physics\ (\bibinfo  {publisher} {Cambridge Univ. Press},\
  \bibinfo {address} {Cambridge, UK},\ \bibinfo {year} {1984})\BibitemShut
  {NoStop}%
\bibitem [{\citenamefont {Boyle}\ and\ \citenamefont
  {Steinhardt}(2008)}]{Boyle:2005se}%
  \BibitemOpen
  \bibfield  {author} {\bibinfo {author} {\bibfnamefont {L.~A.}\ \bibnamefont
  {Boyle}}\ and\ \bibinfo {author} {\bibfnamefont {P.~J.}\ \bibnamefont
  {Steinhardt}},\ }\href {\doibase 10.1103/PhysRevD.77.063504} {\bibfield
  {journal} {\bibinfo  {journal} {Phys. Rev. D}\ }\textbf {\bibinfo {volume}
  {77}},\ \bibinfo {pages} {063504} (\bibinfo {year} {2008})},\ \Eprint
  {http://arxiv.org/abs/astro-ph/0512014} {arXiv:astro-ph/0512014} \BibitemShut
  {NoStop}%
\bibitem [{\citenamefont {Bernal}\ and\ \citenamefont
  {Hajkarim}(2019)}]{Bernal:2019lpc}%
  \BibitemOpen
  \bibfield  {author} {\bibinfo {author} {\bibfnamefont {N.}~\bibnamefont
  {Bernal}}\ and\ \bibinfo {author} {\bibfnamefont {F.}~\bibnamefont
  {Hajkarim}},\ }\href {\doibase 10.1103/PhysRevD.100.063502} {\bibfield
  {journal} {\bibinfo  {journal} {Phys. Rev. D}\ }\textbf {\bibinfo {volume}
  {100}},\ \bibinfo {pages} {063502} (\bibinfo {year} {2019})},\ \Eprint
  {http://arxiv.org/abs/1905.10410} {arXiv:1905.10410 [astro-ph.CO]}
  \BibitemShut {NoStop}%
\bibitem [{\citenamefont {Kite}\ \emph {et~al.}(2021)\citenamefont {Kite},
  \citenamefont {Chluba}, \citenamefont {Ravenni},\ and\ \citenamefont
  {Patil}}]{Kite:2021yoe}%
  \BibitemOpen
  \bibfield  {author} {\bibinfo {author} {\bibfnamefont {T.}~\bibnamefont
  {Kite}}, \bibinfo {author} {\bibfnamefont {J.}~\bibnamefont {Chluba}},
  \bibinfo {author} {\bibfnamefont {A.}~\bibnamefont {Ravenni}}, \ and\
  \bibinfo {author} {\bibfnamefont {S.~P.}\ \bibnamefont {Patil}},\ }\href
  {\doibase 10.1093/mnras/stab3125} {\bibfield  {journal} {\bibinfo  {journal}
  {Mon. Not. Roy. Astron. Soc.}\ }\textbf {\bibinfo {volume} {509}},\ \bibinfo
  {pages} {1366} (\bibinfo {year} {2021})},\ \Eprint
  {http://arxiv.org/abs/2107.13351} {arXiv:2107.13351 [astro-ph.CO]}
  \BibitemShut {NoStop}%
\bibitem [{\citenamefont {Kolb}\ and\ \citenamefont
  {Turner}(1990)}]{Kolb:1990vq}%
  \BibitemOpen
  \bibfield  {author} {\bibinfo {author} {\bibfnamefont {E.~W.}\ \bibnamefont
  {Kolb}}\ and\ \bibinfo {author} {\bibfnamefont {M.~S.}\ \bibnamefont
  {Turner}},\ }\href {\doibase 10.1201/9780429492860} {\emph {\bibinfo {title}
  {{The Early Universe}}}},\ Vol.~\bibinfo {volume} {69}\ (\bibinfo {year}
  {1990})\BibitemShut {NoStop}%
\bibitem [{\citenamefont {Saikawa}\ and\ \citenamefont
  {Shirai}(2018)}]{Saikawa:2018rcs}%
  \BibitemOpen
  \bibfield  {author} {\bibinfo {author} {\bibfnamefont {K.}~\bibnamefont
  {Saikawa}}\ and\ \bibinfo {author} {\bibfnamefont {S.}~\bibnamefont
  {Shirai}},\ }\href {\doibase 10.1088/1475-7516/2018/05/035} {\bibfield
  {journal} {\bibinfo  {journal} {JCAP}\ }\textbf {\bibinfo {volume} {05}},\
  \bibinfo {pages} {035} (\bibinfo {year} {2018})},\ \Eprint
  {http://arxiv.org/abs/1803.01038} {arXiv:1803.01038 [hep-ph]} \BibitemShut
  {NoStop}%
\bibitem [{\citenamefont {Laine}\ and\ \citenamefont
  {Schroder}(2006)}]{Laine:2006cp}%
  \BibitemOpen
  \bibfield  {author} {\bibinfo {author} {\bibfnamefont {M.}~\bibnamefont
  {Laine}}\ and\ \bibinfo {author} {\bibfnamefont {Y.}~\bibnamefont
  {Schroder}},\ }\href {\doibase 10.1103/PhysRevD.73.085009} {\bibfield
  {journal} {\bibinfo  {journal} {Phys. Rev. D}\ }\textbf {\bibinfo {volume}
  {73}},\ \bibinfo {pages} {085009} (\bibinfo {year} {2006})},\ \Eprint
  {http://arxiv.org/abs/hep-ph/0603048} {arXiv:hep-ph/0603048} \BibitemShut
  {NoStop}%
\bibitem [{\citenamefont {Schmitz}(2021)}]{Schmitz:2020syl}%
  \BibitemOpen
  \bibfield  {author} {\bibinfo {author} {\bibfnamefont {K.}~\bibnamefont
  {Schmitz}},\ }\href {\doibase 10.1007/JHEP01(2021)097} {\bibfield  {journal}
  {\bibinfo  {journal} {JHEP}\ }\textbf {\bibinfo {volume} {01}},\ \bibinfo
  {pages} {097} (\bibinfo {year} {2021})},\ \Eprint
  {http://arxiv.org/abs/2002.04615} {arXiv:2002.04615 [hep-ph]} \BibitemShut
  {NoStop}%
\bibitem [{\citenamefont {Thrane}\ and\ \citenamefont
  {Romano}(2013)}]{Thrane:2013}%
  \BibitemOpen
  \bibfield  {author} {\bibinfo {author} {\bibfnamefont {E.}~\bibnamefont
  {Thrane}}\ and\ \bibinfo {author} {\bibfnamefont {J.~D.}\ \bibnamefont
  {Romano}},\ }\href {\doibase 10.1103/PhysRevD.88.124032} {\bibfield
  {journal} {\bibinfo  {journal} {Phys. Rev. D}\ }\textbf {\bibinfo {volume}
  {88}},\ \bibinfo {pages} {124032} (\bibinfo {year} {2013})}\BibitemShut
  {NoStop}%
\bibitem [{\citenamefont {Alonso}\ \emph {et~al.}(2020)\citenamefont {Alonso},
  \citenamefont {Contaldi}, \citenamefont {Cusin}, \citenamefont {Ferreira},\
  and\ \citenamefont {Renzini}}]{Alonso:2020rar}%
  \BibitemOpen
  \bibfield  {author} {\bibinfo {author} {\bibfnamefont {D.}~\bibnamefont
  {Alonso}}, \bibinfo {author} {\bibfnamefont {C.~R.}\ \bibnamefont
  {Contaldi}}, \bibinfo {author} {\bibfnamefont {G.}~\bibnamefont {Cusin}},
  \bibinfo {author} {\bibfnamefont {P.~G.}\ \bibnamefont {Ferreira}}, \ and\
  \bibinfo {author} {\bibfnamefont {A.~I.}\ \bibnamefont {Renzini}},\ }\href
  {\doibase 10.1103/PhysRevD.101.124048} {\bibfield  {journal} {\bibinfo
  {journal} {Phys. Rev. D}\ }\textbf {\bibinfo {volume} {101}},\ \bibinfo
  {pages} {124048} (\bibinfo {year} {2020})},\ \Eprint
  {http://arxiv.org/abs/2005.03001} {arXiv:2005.03001 [astro-ph.CO]}
  \BibitemShut {NoStop}%
\bibitem [{\citenamefont {D'Eramo}\ and\ \citenamefont
  {Schmitz}(2019)}]{DEramo:2019tit}%
  \BibitemOpen
  \bibfield  {author} {\bibinfo {author} {\bibfnamefont {F.}~\bibnamefont
  {D'Eramo}}\ and\ \bibinfo {author} {\bibfnamefont {K.}~\bibnamefont
  {Schmitz}},\ }\href {\doibase 10.1103/PhysRevResearch.1.013010} {\bibfield
  {journal} {\bibinfo  {journal} {Phys. Rev. Research.}\ }\textbf {\bibinfo
  {volume} {1}},\ \bibinfo {pages} {013010} (\bibinfo {year} {2019})},\ \Eprint
  {http://arxiv.org/abs/1904.07870} {arXiv:1904.07870 [hep-ph]} \BibitemShut
  {NoStop}%
\bibitem [{\citenamefont {Campeti}\ \emph {et~al.}(2021)\citenamefont
  {Campeti}, \citenamefont {Komatsu}, \citenamefont {Poletti},\ and\
  \citenamefont {Baccigalupi}}]{Campeti:2020xwn}%
  \BibitemOpen
  \bibfield  {author} {\bibinfo {author} {\bibfnamefont {P.}~\bibnamefont
  {Campeti}}, \bibinfo {author} {\bibfnamefont {E.}~\bibnamefont {Komatsu}},
  \bibinfo {author} {\bibfnamefont {D.}~\bibnamefont {Poletti}}, \ and\
  \bibinfo {author} {\bibfnamefont {C.}~\bibnamefont {Baccigalupi}},\ }\href
  {\doibase 10.1088/1475-7516/2021/01/012} {\bibfield  {journal} {\bibinfo
  {journal} {JCAP}\ }\textbf {\bibinfo {volume} {01}},\ \bibinfo {pages} {012}
  (\bibinfo {year} {2021})},\ \Eprint {http://arxiv.org/abs/2007.04241}
  {arXiv:2007.04241 [astro-ph.CO]} \BibitemShut {NoStop}%
\bibitem [{\citenamefont {Chowdhury}\ \emph {et~al.}(2022)\citenamefont
  {Chowdhury}, \citenamefont {Tasinato},\ and\ \citenamefont
  {Zavala}}]{Chowdhury:2022gdc}%
  \BibitemOpen
  \bibfield  {author} {\bibinfo {author} {\bibfnamefont {D.}~\bibnamefont
  {Chowdhury}}, \bibinfo {author} {\bibfnamefont {G.}~\bibnamefont {Tasinato}},
  \ and\ \bibinfo {author} {\bibfnamefont {I.}~\bibnamefont {Zavala}},\ }\href
  {\doibase 10.1088/1475-7516/2022/08/010} {\bibfield  {journal} {\bibinfo
  {journal} {JCAP}\ }\textbf {\bibinfo {volume} {08}},\ \bibinfo {pages} {010}
  (\bibinfo {year} {2022})},\ \Eprint {http://arxiv.org/abs/2204.10218}
  {arXiv:2204.10218 [gr-qc]} \BibitemShut {NoStop}%
\end{thebibliography}%

\end{document}